# Spin orientations of the spin-half $Ir^{4+}$ ions in $Sr_3NiIrO_6$, $Sr_2IrO_4$ and $Na_2IrO_3$: Density functional, perturbation theory and Madelung potential analyses

Elijah E. Gordon[1], Hongjun Xiang[2,3], Jürgen Köhler[4], and Myung-Hwan Whangbo[1,*]

[1] Department of Chemistry, North Carolina State University, Raleigh, NC 27695-8204, USA

[2] Key Laboratory of Computational Physical Sciences (Ministry of Education), State Key Laboratory of Surface Physics, Collaborative Innovation Center of Advanced Microstructures, and Department of Physics, Fudan University, Shanghai 200433, P. R. China

[3] Collaborative Innovation Center of Advanced Microstructures, Nanjing 210093, P. R. China

[4] Max-Planck-Institut für Festkörperforschung, D-70569 Stuttgart, Germany

Keywords: Magnetic anisotropy, spin-half ions, spin-orbit coupling, Madelung potential

**Abstract**

The spins of the low-spin $Ir^{4+}$ (S = 1/2, $d^5$) ions at the octahedral sites of the oxides $Sr_3NiIrO_6$, $Sr_2IrO_4$ and $Na_2IrO_3$ exhibit preferred orientations with respect to their $IrO_6$ octahedra. We evaluated the magnetic anisotropies of these S = 1/2 ions on the basis of density functional theory (DFT) calculations including spin-orbit coupling (SOC), and probed their origin by performing perturbation theory analyses with SOC as perturbation within the LS coupling scheme. The observed spin orientations of $Sr_3NiIrO_6$ and $Sr_2IrO_4$ are correctly predicted by DFT calculations, and are accounted for by the perturbation theory analysis. As for the spin orientation of $Na_2IrO_3$, both experimental studies and DFT calculations have not been unequivocal. Our analysis reveals that the $Ir^{4+}$ spin orientation of $Na_2IrO_3$ should have nonzero components along the c- and a-axes directions. The spin orientations determined by DFT calculations are sensitive to the accuracy of the crystal structures employed, which is explained by perturbation theory analyses when interactions between adjacent $Ir^{4+}$ ions are taken into consideration. There are indications implying that the 5d electrons of $Na_2IrO_3$ are less strongly localized compared with those of $Sr_3NiIrO_6$ and $Sr_2IrO_4$. This implication was confirmed by showing that the Madelung potentials of the $Ir^{4+}$ ions are less negative in $Na_2IrO_3$ than in $Sr_3NiIrO_6$ and $Sr_2IrO_4$. Most transition-metal S = 1/2 ions do have magnetic anisotropies because the SOC induces interactions among their crystal-field split d-states, and the associated mixing of the states modifies only the orbital parts of the states. This finding cannot be mimicked by a spin Hamiltonian because this model Hamiltonian lacks the orbital degree of freedom, thereby leading to the spin-half syndrome. The spin-orbit entanglement for the 5d spin-half ions $Ir^{4+}$ is not as strong as has been assumed.

**1. Introduction**

Compounds made up of elements with large spin-orbit coupling (SOC) constant $\lambda$ have received much attention recently. They give rise to topological insulators,[1] Rashba-Dresselhaus effects,[2-4] valleytronics,[5,6] and spin-textured bands.[7,8] For magnetic ions in discrete molecules and crystalline solids, an important consequence of SOC is their preferred spin orientations in coordinate space. A uniaxial magnetic ion has a nonzero magnetic moment only in one direction in coordinate space, while an isotropic magnetic ion has a nonzero moment in all directions with equal magnitude. An anisotropic magnetic ion, lying between these two cases, has a moment with magnitude depending on the spin direction. In transition-metal oxides a magnetic ion $M$ forms a $M\text{O}_n$ polyhedron with its surrounding first-coordinate O ligands (typically, n = 4 − 6), so its d-states are split as a result of the $\sigma$- and $\pi$-antibonding interactions of the M d-orbitals largely with the O 2p orbitals.[9] These split d-states are commonly referred to as the crystal-field split d-states. In describing the electronic states of a magnetic system, an electronic Hamiltonian employs both the orbital and spin degrees of freedom. A transition-metal magnetic ion exhibits a preferred spin orientation because SOC induces interactions among the split d-states and because the associated energy lowering depends on the spin orientation with respect to the $M\text{O}_n$ polyhedron.[10-12] For any given magnetic ion, its d-state splitting does not depend on the value of its spin S, and so are the SOC-induced interactions between the split d-states. Therefore, the preferred spin orientation of an S = 1/2 magnetic ion should be governed by SOC, just as are S > 1/2 magnetic ions. This point was recently demonstrated for several S = 1/2 ions by DFT calculations[12] and also by perturbation theory analysis.[12,13] For a long time it has been believed that S = 1/2 ions do not possess magnetic anisotropy. This conceptual impasse, recently termed the "spin-half syndrome",[13] originates from the use of a spin Hamiltonian, which represents the states of a magnetic ion by using only the spin degree of freedom. A magnetic ion with an unevenly-filled degenerate level has an unquenched orbital momentum $\vec{L}$. For a magnetic ion with spin momentum $\vec{S}$ and orbital momentum $\vec{L}$, the total momentum is given by $\vec{J} = \vec{S} + \vec{L}$. "S = 1/2" is not sufficient for a magnetic ion to be magnetically isotropic.[13,14] In most cases, for a S = 1/2 magnetic ion forming bonds with surrounding ligands in a discrete molecule or a crystalline solid, the orbital momentum quenching is not complete, giving rise to a small nonzero orbital momentum $\delta\vec{L}$ on it, so that the SOC between $\vec{S}$ and $\delta\vec{L}$ gives rise to its magnetic anisotropy.[12,13]

The S = 1/2 ions of 5d elements are found for low-spin $Ir^{4+}$ (S = 1/2, $d^5$) ions at octahedral sites, which include $Y_2Ir_2O_7$, $Sr_3NiIrO_6$, $Sr_2IrO_4$, and $Na_2IrO_3$. Each of crystal-field split d-states can be further split by SOC[15-17] thereby introducing a band gap at the Fermi level, in particular, for 5d ions with large SOC constant $\lambda$. That the combination of strong SOC and weak electron correlation creates a magnetic insulating state was first reported for $Ba_2NaOsO_6$.[15] This phenomenon is now considered as a consequence of strong spin-orbit entanglement,[16] and the resulting magnetic insulating state described as a SOC-induced Mott insulating state[17] or spin-orbit Mott insulating state.[18] $Y_2Ir_2O_7$ exhibits spin frustration[19,20] due to the pyrochlore arrangement[21] of its $Ir^{4+}$ ions. $Sr_3NiIrO_6$, $Sr_2IrO_4$ and $Na_2IrO_3$ have no spin frustration. Both $Sr_3NiIrO_6$ and $Sr_2IrO_4$ are magnetic insulators, that is, they have a band gap at all temperature. $Sr_3NiIrO_6$ exhibits uniaxial magnetism[22-24] while $Sr_2IrO_4$ shows an anisotropic magnetism.[25-27] This difference between the two oxides reflects the fact that their $IrO_6$ octahedra have different electron configurations due to the difference in their shapes, as recently pointed out in the perturbation theory analysis[13] in which SOC is taken as perturbation with the crystal-field split d-states of $MO_n$ as unperturbed states. $Na_2IrO_3$ has been thought to be a magnetic insulator,[28,29] but a recent DFT study suggested that it might be a Slater insulator,[30] a system with a partially-filled bands and weak electron correlation.

A Slater insulator opens a band gap when it undergoes a metal-insulator transition at a temperature below which an antiferromagnetic (AFM) ordering sets in.[31] In other words, the d-states of a Slater insulator are less localized than are those of a spin-orbit Mott insulator. It is of interest to examine what structural and electronic features distinguish $Na_2IrO_3$ from $Sr_3NiIrO_6$ and $Sr_2IrO_4$ in this regard.

As for the preferred spin orientation of the $Ir^{4+}$ ions of $Na_2IrO_3$, the experimental studies have not been unequivocal [e.g., the ∥c,[32] ∥a,[33-35] and ∥(a+c) [36] directions], nor have been the DFT studies [e.g., ∥b,[32] ∥(0.12a + 0.32c*),[30] and ∥(a+c) [37] directions]. So far, the cause for these controversial observations has not been understood. Nevertheless, we notice that the DFT studies employed different crystal structures of $Na_2IrO_3$, and the fractional coordinates of the oxygen atoms in these crystal structures are poorly determined (with standard deviations in the third decimal places). This implies that the use of the crystal structures with different accuracies might have caused the controversial results reported. It is important to explore this possibility.

In predicting the preferred spin orientations of magnetic ions $M$ in magnetic oxides in terms of the SOC-induced HOMO-LUMO interactions, one needs the split d-states of their local $MO_n$ polyhedra.[13] For simplicity, these can be deduced by considering an isolated $MO_n$ polyhedron, which implicitly assumes that the split d-states thus obtained remain unchanged by the interactions between adjacent $M$ ions (hereafter the intersite interactions). In general, this assumption is reasonable for 3d ions because the $M$ 3d and O 2p orbitals do not differ strongly in their contractedness so that the associated crystal-field splitting of an isolated $MO_n$ polyhedron is strong. However, the $M$ 5d orbitals are much more diffuse than O 2p orbitals thereby weakening the d-state splitting of an isolated $MO_n$ polyhedron. For oxides of 5d ions, therefore, the relative ordering of the split d-states deduced from an isolated $MO_n$ polyhedron might be altered by the inter-site interaction. It is of interest to examine how the intersite interactions affect the split d-states of the $Ir^{4+}$ ions in $Sr_3NiIrO_6$, $Sr_2IrO_4$ and $Na_2IrO_3$ and hence influence their spin orientations.

In the present work we investigate the questions raised above on the basis of DFT, perturbation theory, and Madelung potential analyses. Our work is organized as follows: Section 2 describes the important structural features of $Sr_3NiIrO_6$, $Sr_2IrO_4$, and $Na_2IrO_3$ that are needed in our subsequent discussions. Section 3 briefly reviews how to use the SOC-induced HOMO-LUMO interactions in predicting the preferred spin orientations of magnetic ions.[13] In Section 4 we quantitatively evaluate the observed spin orientations of the $Ir^{4+}$ ions in $Sr_3NiIrO_6$, $Sr_2IrO_4$, and $Na_2IrO_3$ on the basis of DFT calculations, and subsequently examine the results in terms of perturbation theory analysis including the intersite interactions. In Section 5 we examine if the Madelung potentials (i.e., the electrostatic potentials) acting on the $Ir^{4+}$ ions of $Sr_3NiIrO_6$, $Sr_2IrO_4$, and $Na_2IrO_3$ affect the extent of electron localization of their 5d electrons. The origin of the spin-half syndrome is discussed in Section 6. In Section 7 we discuss two conceptual issues arising from the strong SOC constant $\lambda$ of $Ir^{4+}$ ions, namely, the consideration of SOC as perturbation and the comparison of the LS vs. jj coupling schemes for the 5d spin-half ions $Ir^{4+}$. Our conclusions are briefly summarized in Section 8.

## 2. Structural features and split d-states

$Sr_3NiIrO_6$ consists of $NiIrO_6$ chains in which $IrO_6$ octahedra alternate with the $NiO_6$ trigonal prisms by sharing their triangular faces (**Fig. 1a**).[38] These $NiIrO_6$ chains, aligned along the c-axis and separated by $Sr^{2+}$ ions, have a trigonal arrangement (**Fig. 1b**). $Sr_3NiIrO_6$ exhibits a uniaxial magnetism with the magnetic moment aligned parallel to the chain direction,[22,24] and the spins of adjacent $Ni^{2+}$ (S = 1, $d^8$) and $Ir^{4+}$ (S = 1/2, $d^5$) ions in each $NiIrO_6$ chain have an AFM coupling.[22] Each $IrO_6$ octahedron is nearly regular in shape with its 3-fold-rotational axis aligned along the c-direction (i.e., the chain direction).

In the layered oxide $Sr_2IrO_4$, axially-elongated $IrO_6$ octahedra share the oxygen atoms of their equatorial $Ir-O_{eq}$ bonds to form $IrO_4$ layers with bent $Ir-O_{eq}-Ir$ linkages (i.e., $\angle Ir-O_{eq}-Ir = 157.20°$) (**Fig. 1c**).[39] A unit cell has four such layers, and the elongated $Ir-O_{ax}$ bonds of each $IrO_6$ octahedron deviates slightly from the c-axis direction because it is slightly tilted with respect to the ab-plane of the layer.[39] Neutron diffraction studies show that the spin of each $Ir^{4+}$ ion is perpendicular to the elongated $Ir-O_{ax}$ bonds, and the adjacent $Ir^{4+}$ spins in each layer have an AFM coupling while this intra-layer spin arrangement have a ferromagnetic (FM) coupling along the c-direction (**Fig. 1d**).[25,26]

$Na_2IrO_3$ consists of honeycomb layers made up of edge-sharing $IrO_6$ octahedra with the center of each $Ir_6$ hexagon occupied by a $Na^+$ ion (**Fig. 1e**), and such $NaIrO_3$ layers alternate with layers of $Na^+$ cations (**Fig. 1f**).[32-34,40] $Na_2IrO_3$ was initially thought to crystallize in a C2/c structure,[32,40] but was later found to have a C2/m structure.[33,34] There are two known C2/m structures, one determined at 300 K[33] and the other at 125 K.[34] In the crystal structure determined at 125 K,[34] each $IrO_6$ octahedron is substantially compressed along the direction perpendicular to the layer, i.e., the c*-direction (i.e., perpendicular to the ab-plane and slightly away from the c-direction, see **Fig. 1f**). The magnetic susceptibility study on a single crystal sample show that the susceptibility is greater at all temperatures examined (below 400 K) when the probe magnetic field is applied along the c-direction than along any other direction.[32] This suggests that the preferred spin orientation is close to the ∥c-direction. The ∥a spin-direction was reported in neutron diffraction[34] and neutron scattering[33] studies, and nearly the ∥a [35] and the ∥(a+c) spin-directions [36] in resonant x-ray magnetic scattering studies. In terms of DFT studies, the ∥b direction is the preferred spin direction when the C2/c structure is used,[32] while the ∥(a+c) [37] and the ∥(0.12a + 0.32c*) [30] spin-directions were predicted using the room-temperature C2/m structure.[33]

For a magnetic ion $M$ present in a magnetic insulating oxide, one might approximate the crystal-field split d-states of its $MO_n$ polyhedra by those for an isolated $MO_n$ polyhedron. In spin-polarized electronic-structure description, these d-states are split into the up-spin and down-spin states, with the up-spin d-states lying below down-spin d-states by convention. Consider first the $IrO_6$ octahedra of $Sr_3NiIrO_6$. With the local z-axis taken along the 3-fold-rotational axis, the $t_{2g}$ state of each $IrO_6$ octahedron is divided into $1a$, $1e_x$ and $1e_y$ states (**Fig. 2a**).[13] In the spin-polarized electronic-structure description, the up-spin and down-spin $t_{2g}$ states are split as in **Fig. 2b** and **2c**. These two electron configurations differ in how the three down-spin states are occupied by two electrons.

With the local z-axis of an axially-elongated $IrO_6$ octahedron of $Sr_2IrO_4$ taken along the Ir-$O_{ax}$ bonds, the $t_{2g}$ state is split into (xz, yz) < xy. Thus, in the spin-polarized electronic-structure description, the up-spin and down-spin $t_{2g}$ states are occupied as in **Fig. 3a**. (For the simplicity of presentation, the split between the up-spin and down-spin states is exaggerated here and hereafter. This simplification does not affect our discussion, because the HOMO and LUMO occur in the down-spin states.) However, if $Sr_2IrO_4$ were composed of axially-compressed $IrO_6$ octahedra, then the $t_{2g}$ state would be split into xy < (xz, yz) so that the up-spin and down-spin $t_{2g}$ states would be occupied as in **Fig. 3b**.

Strictly speaking, each $IrO_6$ octahedron of $Na_2IrO_3$ has no 3-fold-rotational symmetry but has a pseudo 3-fold rotational axis along the direction perpendicular to the layer (i.e., along the c*-direction, **Fig. 1f**). Due to the compression of the $IrO_6$ octahedron along this axis, its $t_{2g}$ state is split into 1a < (1e$_x$, 1e$_y$), where 1e$_x$ and 1e$_y$ are approximately degenerate, so that the t$_{2g}$ state would be occupied as depicted in **Fig. 3c**.

## 3. Perturbation theory analysis of preferred spin orientation

The preferred spin direction of a magnetic ion $M$ in a magnetic insulator can be parallel to the local z-axis of its $M$O$_n$ polyhedron (the ∥z-direction, which is along the axis of the highest rotational symmetry), perpendicular to it (the ⊥z-direction) or a certain direction in between the two. Quantitatively, the preferred spin orientation is the direction of the spin for which the total energy of a system under consideration is minimum on the basis of DFT calculations (see Section 4). The objective of this section is to discuss the simple rules based on perturbation theory analysis,[13] with which one can predict and/or analyze the outcome of such calculations as well as experimental results.

If we use two independent coordinates for the orbital $\hat{L}$ and the spin $\hat{S}$, for example, the (x, y, z) coordinate for $\hat{L}$ and the (x′, y′, z′) coordinate for $\hat{S}$, the SOC Hamiltonian $\hat{H}_{SO} = \lambda\hat{S}\cdot\hat{L}$ is rewritten as $\hat{H}_{SO} = \hat{H}_{SO}^0 + \hat{H}_{SO}'$,[10-13,41] where

$$\hat{H}_{SO}^0 = \lambda\hat{S}_{z'}\left(\hat{L}_z\cos\theta + \frac{1}{2}\hat{L}_+e^{-i\phi}\sin\theta + \frac{1}{2}\hat{L}_-e^{+i\phi}\sin\theta\right) \tag{1a}$$

$$= \lambda\hat{S}_{z'}(\hat{L}_z\cos\theta + \hat{L}_x\sin\theta\cos\phi + \hat{L}_y\sin\theta\sin\phi)\,. \tag{1b}$$

$$\hat{H}_{SO}' = \frac{\lambda}{2}(\hat{S}_{+'} + \hat{S}_{-'})\left(-\hat{L}_z\sin\theta + \hat{L}_x\cos\theta\cos\phi + \hat{L}_y\cos\theta\sin\phi\right) \tag{2}$$

The SOC-induced energy lowering occurs from the interactions of the occupied states, $\psi_o$, with unoccupied states, $\psi_u$, of a $M$O$_n$ polyhedron, and is governed by the interaction matrix elements $\langle\psi_o|\hat{H}_{SO}|\psi_u\rangle$. Whether $\langle\psi_o|\hat{H}_{SO}|\psi_u\rangle$ is nonzero or not is determined by the spin states of $\psi_o$ and

$\psi_u$ and also by the minimum difference $\left|\Delta L_z\right|$ between the magnetic quantum numbers $L_z$ of the d-orbitals constituting $\psi_o$ and $\psi_u$. When both $\psi_o$ and $\psi_u$ are up-spin states, or down-spin states, the term $\left\langle \psi_o \left| \hat{H}_{SO} \right| \psi_u \right\rangle$ is determined by $\left\langle \psi_o \left| \hat{H}_{SO}^0 \right| \psi_u \right\rangle$ in Eq. 1. When $\psi_o$ and $\psi_u$ are opposite in spin states, the term $\left\langle \psi_o \left| \hat{H}_{SO} \right| \psi_u \right\rangle$ is determined by $\left\langle \psi_o \left| \hat{H}_{SO}' \right| \psi_u \right\rangle$ in Eq. 2. Note that $L_z = 0$ for $3z^2$-$r^2$, $L_z = \pm 1$ for $\{xz, yz\}$, and $L_z = \pm 2$ for $\{xy, x^2\text{-}y^2\}$. Thus,

$\left|\Delta L_z\right| = 0$ between xz and yz, and between xy and $x^2$-$y^2$

$\left|\Delta L_z\right| = 1$ between $3z^2$-$r^2$ and $\{xz, yz\}$, and between $\{xz, yz\}$ and $\{xy, x^2\text{-}y^2\}$.

$\left|\Delta L_z\right| = 2$ between $3z^2$-$r^2$ and $\{xy, x^2\text{-}y^2\}$

Therefore, we arrive at the following rules for predicting the preferred spin orientations:

Between d-states of a same spin:

$$\left\langle \psi_o \left| \hat{H}_{SO}^0 \right| \psi_u \right\rangle \propto \begin{cases} \cos\theta, & \text{if } \left|\Delta L_z\right| = 0 \\ \sin\theta, & \text{if } \left|\Delta L_z\right| = 1 \end{cases} \qquad (3)$$

Between d-states of opposite spins:

$$\left\langle \psi_o \left| \hat{H}_{SO}' \right| \psi_u \right\rangle \propto \begin{cases} \sin\theta, & \text{if } \left|\Delta L_z\right| = 0 \\ \cos\theta, & \text{if } \left|\Delta L_z\right| = 1 \end{cases} \qquad (4)$$

For SOC-induced interactions between same-spin states, $\left\langle \psi_o \left| \hat{H}_{SO}^0 \right| \psi_u \right\rangle$ is maximum at $\theta = 0°$ (the $\|z$ spin orientation) if $\left|\Delta L_z\right| = 0$, but is maximum at $\theta = 90°$ if $\left|\Delta L_z\right| = 1$ (the $\perp z$ spin orientation). For SOC-induced interactions between opposite-spin states, the exactly opposite spin orientations are predicted. Under SOC, $\psi_o$ and $\psi_u$ do not interact when $\left|\Delta L_z\right| > 1$, because $\left\langle \psi_o \left| \hat{H}_{SO} \right| \psi_u \right\rangle = 0$ in such a case. In applying these rules to predict the preferred spin orientation of a magnetic ion $M$, we note that the most important interaction is the one between the HOMO and LUMO of its $M$O$_n$

polyhedron. According to Eq. 3 or 4, the preferred spin orientation is either ∥z or ⊥. For the preferred spin orientation to lie in between the ∥z and ⊥z directions, therefore, there must be two "HOMO-LUMO" interactions that predict different spin orientations (one for ∥z, and the other ⊥z). Such a situation occurs for $Na_2IrO_3$, as will be discussed below.

For the $Ir^{4+}$ ions of $Sr_3NiIrO_6$ in which adjacent $Ni^{2+}$ (S = 1) and $Ir^{4+}$ (S = 1/2) spins have an AFM coupling in each $NiIrO_6$ chain (see Section 4.2), the HOMO and LUMO occur from the down-spin electron configuration $(1a\downarrow)^1(1e_x\downarrow, 1e_y\downarrow)^1$ (**Fig. 2b**), so the preferred spin orientation is the ∥z direction because $\left|\Delta L_z\right| = 0$. (By using Eq. 1b and Table 3 of ref. 10, it can be readily shown that

$$\left\langle 1e_x \left| \hat{L}_z \right| 1e_y \right\rangle = \tfrac{2}{3}\left\langle xy \left| \hat{L}_z \right| x^2 - y^2 \right\rangle + \tfrac{1}{3}\left\langle xz \left| \hat{L}_z \right| yz \right\rangle = \tfrac{4}{3}i - \tfrac{1}{3}i = i\,,$$

$$\left\langle 1e_x \left| \hat{L}_x \right| 1e_y \right\rangle = -\tfrac{\sqrt{2}}{3}\left\langle xy \left| \hat{L}_x \right| yz \right\rangle - \tfrac{\sqrt{2}}{3}\left\langle xz \left| \hat{L}_x \right| x^2 - y^2 \right\rangle = -\tfrac{\sqrt{2}}{3}(0 + 0) = 0\,,$$

$$\left\langle 1e_x \left| \hat{L}_y \right| 1e_y \right\rangle = -\tfrac{\sqrt{2}}{3}\left\langle xy \left| \hat{L}_y \right| yz \right\rangle - \tfrac{\sqrt{2}}{3}\left\langle xz \left| \hat{L}_y \right| x^2 - y^2 \right\rangle = -\tfrac{\sqrt{2}}{3}(i - i) = 0\,.$$

Thus, $\left|\Delta L_z\right| = 0$.) This is in agreement with experiment.[22,24] However, if adjacent $Ni^{2+}$ (S = 1) and $Ir^{4+}$ (S = 1/2) spins are forced to have an FM coupling in each $NiIrO_6$ chain (see Section 4.2), the HOMO and LUMO occur from the down-spin electron configuration $(1a\downarrow)^0(1e_x\downarrow, 1e_y\downarrow)^2$ (**Fig. 2c**) so that the preferred spin orientation is the ⊥z direction because $\left|\Delta L_z\right| = 1$.

For the axially-elongated $IrO_6$ octahedron of $Sr_2IrO_4$, the $Ir^{4+}$ ion of an isolated $IrO_6$ octahedron has the down-spin electron configuration $(xz\downarrow, yz\downarrow)^2(xy\downarrow)^0$ (**Fig. 3a**), from which the HOMO and LUMO are found. Thus, the preferred spin orientation is the ⊥z direction because $\left|\Delta L_z\right| = 1$. This is in agreement with experiment.[25,26] If $Sr_2IrO_4$ were made up of axially-compressed $IrO_6$ octahedra so that the $Ir^{4+}$ ion has the down-spin electron configuration $(xy\downarrow)^1(xz\downarrow, yz\downarrow)^1$ (**Fig. 3b**), then the preferred spin orientation of $Ir^{4+}$ would be the $\|z$ direction because $\left|\Delta L_z\right| = 0$. However, a recent ESR study[42] of $Sr_2IrO_4$ showed that the ordering of the split $t_{2g}$-states should be $xy < (xz, yz)$ rather than $(xz, yz) < xy$, although it consists of axially-elongated $IrO_6$ octahedra. In the following we show that the apparent orbital sequence, $xy < (xz, yz)$, is caused by the intersite interactions (see Section 4.3).

For the $Ir^{4+}$ ion of $Na_2IrO_3$, the HOMO and LUMO occur from the down-spin electron configuration close to $(1a\downarrow)^1(1e_x\downarrow, 1e_y\downarrow)^1$ (**Fig. 3c**), so the preferred spin orientation would be the $\|z$ direction (namely, the c*-direction in **Fig. 1f**) because $\left|\Delta L_z\right| = 0$. Since the c*-direction is not far from the c-direction, this prediction is consistent with the result of the single-crystal magnetic susceptibility measurements.[32] However, the $(1a\downarrow)^1(1e_x\downarrow, 1e_y\downarrow)^1$ configuration, deduced from an isolated $IrO_6$ octahedron, cannot explain the presence of the a-axis component in the observed spin moment.[33-36] The perturbation theory analysis requires the split d-states of an $IrO_6$ octahedron present in $Na_2IrO_3$, not an isolated $IrO_6$ octahedron. The latter does not have the effect of the intersite interactions. If the intersite interactions effectively reduce the energy gap between $1a\downarrow$ and $(1e_x\downarrow, 1e_y\downarrow)$, then the $(1a\downarrow)^0(1e_x\downarrow, 1e_y\downarrow)^2$ configuration ($\left|\Delta L_z\right| = 1$) would also participate in controlling the spin orientation thereby giving rise to the a-axis component. Thus, it is of interest to probe how the relative ordering of the crystal-field split d-states of an isolated $IrO_6$ octahedron might be modified when the intersite interactions between adjacent $Ir^{4+}$ sites are incorporated. We discuss this question in the next section.

## 4. Quantitative evaluation of preferred spin orientation and analysis of intersite effects

### 4.1. DFT calculations

We carry out spin-polarized DFT calculations for $Sr_3NiIrO_6$, $Sr_2IrO_4$ and $Na_2IrO_3$ by employing the projector augmented wave method encoded in the Vienna ab initio simulation package[43] and the generalized gradient approximation of Perdew, Burke and Ernzerhof for the exchange-correlation functionals.[44] The plane wave cutoff energies of 520 eV and the threshold of self-consistent-field energy convergence of $10^{-6}$ eV were used. We employed the DFT plus on-site repulsion U (DFT+U) method[45] to describe the electron correlation associated with the 3d states of Ni and the 5d states of Ir using the effective on-site repulsion $U_{eff} = U - J$ on Ni and on Ir. The preferred spin orientations were determined by performing DFT+U calculations including SOC effects.[46] An alternative way of describing electron correlation is the DFT+hybrid method,[47] in which the mixing coefficient $\alpha$ is the adjustable parameter, which plays the role of $U_{eff}$ in the DFT+U method.

## 4.2. $Sr_3NiIrO_6$

In our DFT+U and DFT+U+SOC calculations for $Sr_3NiIrO_6$, the irreducible Brillouin zone was sampled by a set of $3\times3\times2$ k-points. Since the 5d orbitals are more diffuse than the 3d orbitals, $U_{eff}$ for Ir should be considerably smaller than $U_{eff}$ for Ni. We employed $U_{eff} = 4 - 6$ eV for Ni, typical values for late 3d elements. With $U_{eff} = 4 - 6$ eV for Ni, magnetic insulating states are obtained by using $U_{eff} \geq \sim 2$ eV for Ir, but metallic states by using $U_{eff} = 0$ and 1 eV for Ir. Such metallic states predicted from DFT+U calculations can become magnetic insulating states once SOC is included as long as the $U_{eff}$ is not close to 0 for Ir, as summarized in **Table S1** of the Supplementary Material.[48] In our calculations for $Sr_3NiIrO_6$, we have not encountered the computational problems reported by Sarkar et al. (namely, they failed to stabilize any magnetic configuration other than FM arrangement between Ir and Ni spins, even with application of U.)[49]

It has been shown by Zhang et al.[17] as well as by Ou and Wu[50] that the magnetic insulating state of $Sr_3NiIrO_6$ is reproduced by DFT+U+SOC calculations only when adjacent $Ni^{2+}$ and $Ir^{4+}$ spins have an AFM coupling in each $NiIrO_6$ chain. These studies did not examine the preferred orientations of the $Ni^{2+}$ and $Ir^{4+}$ spins, but it is known experimentally[22,23] that the ||c-direction is their preferred orientation. Sarkar et al. reported the ⊥c-direction as the preferred orientation for the spin arrangement in which adjacent $Ni^{2+}$ and $Ir^{4+}$ spins have an FM coupling in each $NiIrO_6$ chain.[49] These observations suggest that the preferred spin orientations of the $Ni^{2+}$ and $Ir^{4+}$ ions in each $NiIrO_6$ chain depend on the nature of the nearest-neighbor spin exchange coupling. In the following we verify this suggestion and explore its cause.

Our DFT+U calculations show that the AFM arrangement between adjacent $Ni^{2+}$ and $Ir^{4+}$ ions in each $NiIr_6$ chain is substantially more stable than the FM arrangement (**Table 1**) in agreement with the observed AFM arrangement of the adjacent $Ni^{2+}$ and $Ir^{4+}$ spins in each $NiIrO_6$ chain of $Sr_3NiIrO_6$.[22] For the FM and AFM arrangements between adjacent $Ni^{2+}$ and $Ir^{4+}$ ions in each $NiIrO_6$ chain of $Sr_3NiIrO_6$, we examine their magnetic properties on the basis of DFT+U and DFT+U+SOC calculations by varying $U_{eff}$ = 4 – 6 eV for Ni with $U_{eff}$ fixed at 2 eV for Ir. In the case of FM coupling, the ⊥c spin direction is more stable than the ||c spin direction for both $Ni^{2+}$ and $Ir^{4+}$ ions (**Table 2a**). In the case of AFM coupling, however, the $Ni^{2+}$ and $Ir^{4+}$ spins strongly favor the ||c over the ⊥c direction (**Table 2a**). This result is in good agreement with the analysis of the spin wave excitations in $Sr_3NiIrO_6$,[23] which found the ||c orientation to be more stable by 7.2 meV per formula unit (FU). To understand why the preferred spin orientation of the $Ir^{4+}$ ion depends on the nature of spin exchange coupling between adjacent $Ni^{2+}$ and $Ir^{4+}$ spins in each $NiIrO_6$ chain, we calculate the plots of the projected density of states (PDOS) for $Sr_3NiIrO_6$ by DFT+U calculations. The PDOS plots obtained for the AFM and FM arrangements are presented in **Fig. 2d** and **2e**, respectively, where only the down-spin states are shown for simplicity. The PDOS plot for the AFM arrangement is consistent with the local electron configuration $(1a\downarrow)^1(1e_x\downarrow, 1e_y\downarrow)^1$ (**Fig. 2b**), predicting the $\|z$ spin orientation. The PDOS plot for the FM arrangement is consistent with the local configuration $(1a\downarrow)^0(1e_x\downarrow, 1e_y\downarrow)^1$ (**Fig. 2c**), predicting the $\perp z$ spin orientation. The down-spin 1a state lies high in energy for the FM arrangement and becomes unoccupied, because the FM arrangement makes the $3z^2\text{-}r^2$ orbitals of adjacent $Ni^{2+}$ and $Ir^{4+}$ ions interact more strongly than does the AFM arrangement.[10,11,51]

$Sr_3NiPtO_6$ is isostructural with $Sr_3NiIrO_6$ with magnetic $Ni^{2+}$ ($d^8$, S = 1) ions at the trigonal prism sites and nonmagnetic $Pt^{4+}$ (S = 0, $d^6$) ions at the octahedral sites.[52] Our DFT+U+SOC calculations for $Sr_3NiPtO_6$ reveal that the $Ni^{2+}$ spins weakly favor the $\perp c$ direction over the $\|c$ direction (by 0.15, 0.10 and 0.07 meV/FU using $U_{eff}$ = 4, 5 and 6 eV for Ni, respectively). This result is in agreement with experiment.[24] Therefore, the strong preference for the $\|c$ orientation of the $Ni^{2+}$ spins found for $Sr_3NiIrO_6$ implies that the $Ir^{4+}$ spins strongly favors the $\|c$ direction, and that the strong AFM interaction between adjacent $Ni^{2+}$ and $Ir^{4+}$ spins makes the $Ni^{2+}$ spins aligned along the c-direction. To confirm this implication, we calculate the orbital and spin moments ($\mu_L$ and $\mu_S$, respectively) for the $Ni^{2+}$ and $Ir^{4+}$ cations with their spins aligned along the $\|c$ direction. For the FM coupling between adjacent $Ni^{2+}$ and $Ir^{4+}$ spins, the $\mu_L$ of $Ir^{4+}$ is as small in magnitude as that of $Ni^{2+}$ (**Table 2b**). For the AFM coupling, however, the $\mu_L$ of $Ir^{4+}$ is large and is about half the value of its $\mu_S$ (**Table 2b**). This supports the presence of uniaxial magnetism for the $Ir^{4+}$ ions of $Sr_3NiIrO_6$ when adjacent $Ni^{2+}$ and $Ir^{4+}$ spins have an AFM arrangement in each $NiIrO_6$ chain. The total moment $\mu_T = \mu_L + \mu_S$ is ~0.9 $\mu_B$ for $Ir^{4+}$, and ~1.7 $\mu_B$ for $Ni^{2+}$, in good agreement with the neutron diffraction study of Lefrançois et al.[22]

For the case of AFM coupling between adjacent $Ni^{2+}$ and $Ir^{4+}$ spins, the local electron configuration $(1a\downarrow)^1(1e_x\downarrow, 1e_y\downarrow)^1$ of $Ir^{4+}$ causes uniaxial magnetism along the 3-fold-rotational axis. Since this configuration has an unevenly-filled degenerate state, it has an unquenched orbital moment $\vec{L}$ so that SOC gives rise to the total moment $\vec{J} = \vec{S} + \vec{L}$. In the spin-orbit coupled state of the $Ir^{4+}$ ion, the total angular momentum quantum number for the singly-occupied doublet state is given by J = 3/2.[13] Since J > 1/2, the $Ir^{4+}$ ion exhibits a uniaxial magnetism.[13,14]

### 4.3. $Sr_2IrO_4$

Recently, Liu et al.[27] have reported a thorough study on the magnetic anisotropy of the $Ir^{4+}$ ions in $Sr_2IrO_4$ as well as the origin of the spin canting in $Sr_2IrO_4$ by employing DFT+U+SOC calculations. Their calculations showed that the $Ir^{4+}$ ions prefer the $\perp c$ direction even when their $IrO_6$ octahedra are regular in shape so that the $t_{2g}$ states of an isolated $IrO_6$ octahedron remain triply degenerate. Bogdanov et al.[42] evaluated the g-factors of the $Ir^{4+}$ ion in $Sr_2IrO_4$ along the $\|c$ and $\perp c$ directions by performing configuration interaction calculations for a cluster containing an $IrO_6$ octahedron, to confirm the conclusion of their ESR study that the $t_{2g}$-states split as xy < (xz, yz) rather than as (xz, yz) < xy. The aforementioned observations are apparently puzzling from the viewpoints of the split $t_{2g}$ states of an isolated $IrO_6$ octahedron. In this section we first verify the qualitative predictions of the perturbation theory analysis of Section 3, namely, that the magnetic anisotropy of the $Ir^{4+}$ ions is governed by the relative ordering of the split d-states of their $IrO_6$ octahedra, which in turn is determined by the nature of distortion in the $IrO_6$ octahedra. Then, we show that the split d-state patterns of $Sr_2IrO_4$ become different from those of an isolated $IrO_6$ octahedron because of the intersite interactions.

In our DFT+U and DFT+U+SOC calculations for $Sr_2IrO_4$, we employed $U_{eff}$ = 2 eV and sampled the irreducible Brillouin zone by a set of 4×4×1 k-points. To determine the preferred spin orientations of the $Ir^{4+}$ ions in $Sr_2IrO_4$, we carry out DFT+U+SOC calculations with $U_{eff}$ = 2 eV for the observed magnetic structure (namely, the intra-layer AFM arrangement and the inter-layer FM arrangement). As summarized in **Table 3a**, the ⊥c spin orientation (i.e., the ∥a or ∥b orientation) is more stable than the ∥c orientation for the case when the $IrO_6$ octahedra are axially-elongated in agreement with experiment.[25,26] We construct a hypothetical structure of $Sr_2IrO_4$ made up of axially compressed $IrO_6$ octahedra so that their $t_{2g}$ levels are split as xy < (xz, yz) (**Fig. 3b**). For this purpose, we construct a new orthorhombic cell with the cell parameters $c' = 0.9c$, $a' = \sqrt{10/9}\ a$, and $b' = \sqrt{10/9}\ b$. In this compressed structure, which has the same cell volume as does the experimental structure, $Ir-O_{ax}$ = 1.855 Å and $Ir-O_{eq}$ = 2.068 Å. Our DFT+U+SOC calculations show that, in this axially-compressed structure, the preferred spin orientation is the ∥c direction, as expected from the perturbation theory analysis. In addition, the singly-occupied degenerate level gives rise to the electron configuration $(xz, yz)^1$, for which L = 1 so that J = L + S = 3/2. Since J > 1/2, the spin-half $Ir^{4+}$ ion at the axially-compressed octahedral site should be uniaxial. In agreement with this assessment, the $\mu_S$ and $\mu_L$ are the largest for the ∥c spin orientation of the axially-compressed structure (**Table 3b**).

The PDOS plots calculated for the Ir 5d states (around the Fermi level) of the axially-elongated and axially-compressed structures of $Sr_2IrO_4$ are presented in **Fig. 4a** and **4b**, respectively. These patterns differ from the crystal-field split $t_{2g}$ states expected for an isolated $IrO_6$ octahedron shown in **Fig. 3a** and **3b**, respectively. This is due to the effect of the intersite interactions. The bent $Ir-O_{eq}$-Ir linkages in each $IrO_4$ layer do not weaken the π-antibonding interactions between adjacent xz/yz orbitals (**Fig. 4c**), but weaken those between adjacent xy orbitals (**Fig. 4d**). In other words, the π-type interactions between adjacent xz/yz orbitals are stronger than those between adjacent xy orbitals. To a first approximation, the split d-states of an $IrO_6$ octahedron embedded in $Sr_2IrO_4$ and hence having the intersite interactions can be approximated by those of a dimer made up of two adjacent corner-sharing $IrO_6$ octahedra. Then, for the axially-elongated case, the interactions between two adjacent $Ir^{4+}$ sites alter the crystal-field split $t_{2g}$ states as depicted in **Fig. 4e** so that the HOMO has the xy character, and the LUMO the xz/yz character. The latter is consistent with the PDOS plots of **Fig. 4a**, and predicts the ⊥c spin orientation as does the crystal-field split $t_{2g}$ states of an isolated $IrO_6$ octahedron (**Fig. 3a**). This explains why the ESR study of $Sr_2IrO_4$ by Bogdanov et al.[42] is explained by the d-state ordering xy < (xz, yz), despite that it consists of axially-elongated $IrO_6$ octahedra. Furthermore, it is clear from the above discussion that, even for the structure of $Sr_2IrO_4$ made up of regular $IrO_6$ octahedra, the HOMO has the xy character while the LUMO the xz/yz character, thereby explaining the ⊥c spin orientation for this structure.[27]

For the axially-compressed case, the interaction between the xz/yz states of adjacent $Ir^{4+}$ sites is weakened due to the lengthened $Ir-O_{eq}-Ir$ linkage under compression, so that the interactions between two adjacent $Ir^{4+}$ sites alter the crystal-field split $t_{2g}$ states as depicted in **Fig. 4f**. Thus the HOMO and LUMO both have the xy character. The latter is consistent with the PDOS plots of **Fig. 4b** and predicts the ∥c spin orientation, as does the crystal-field split $t_{2g}$ states of an isolated $IrO_6$ octahedron (**Fig. 3b**).

### 4.4. $Na_2IrO_3$

In our DFT+U+SOC calculations for the spin orientation of $Na_2IrO_3$, we employed a (a, b, 2c) supercell using the experimental magnetic structure (except for the spin orientation), a set of $4\times2\times2$ k-points for the irreducible Brillouin zone, and $U_{eff} = 1$ eV. The latter is based on the finding of ref. 36, in which $U_{eff} = 1.1$ eV is used. The need to use a smaller $U_{eff}$ for $Na_2IrO_3$ than for $Sr_3NiIrO_6$ and $Sr_2IrO_4$ (1 vs. 2 eV) leads us to speculate if the 5d electrons are less localized in $Na_2IrO_3$ than in $Sr_3NiIrO_6$ and $Sr_2IrO_4$. This point is consistent with the finding of the DFT+hybrid calculations for $Na_2IrO_3$ that a band gap is obtained using a very small mixing parameter $\alpha = 0.05$.[30] (Typically, $\alpha = 0.2$ is employed.) This point will be discussed further in Section 5.

For the crystal structure of $Na_2IrO_3$, we employed the experimental C2/m structures determined at 125 and 300 K. In addition, noting that the accuracy of their atom positions is rather low (with standard deviations in the third decimal places), we optimized the atom positions with DFT+U calculations for their FM states by relaxing the atom positions until the force variation at each atom is less than 0.02 eV/Å (with the unit cell parameters fixed at the experimental values). We found that the optimization of the atom positions for the 125 K structure keeps the C2/m structure, whereas that for the 300 K leads to a $P\bar{1}$ structure. The optimized atom positions of the 125 and 300 K structures are summarized in **Tables S2** and **S3** of the Supplementary Material.[48]

Our DFT+U+SOC calculations were carried out in two ways; in the "biased" method, we first perform DFT+U calculations with high $U_{eff}$ (e.g., 4 eV) to obtain the charge density that produces the observed magnetic structure and then use the resulting density for subsequent DFT+U and DFT+U+SOC calculations using lower $U_{eff}$. In the unbiased method, no such biased charge density was used to begin the calculations.

We first examine our results obtained by using the experimental C2/m structure determined at 125 K. For $U_{eff} \approx 2$ eV or smaller, the unbiased DFT+U calculations predict no band gap but the biased DFT+U+SOC calculations do. The biased DFT+U calculations give a band gap if $U_{eff} \approx 2$ eV, but do not if $U_{eff} \approx 1$ eV or smaller. As for the preferred spin orientation, the unbiased

DFT+U+SOC calculations converge approximately to the ∥(2a + c*) direction if the starting spin orientation is either ∥a or ∥c, but to the ∥b direction if the starting spin orientation is ∥b. Furthermore, the ∥b orientation is more stable than the ∥(2a + c*) direction by 1.72 meV/Ir. (Note that our DFT+U+SOC calculations were performed for supercell containing eight FUs.) In contrast, the biased DFT+U+SOC calculations converge approximately to the ∥(2a + c*) direction, regardless of whether the starting spin orientation is ∥a, ∥b or ∥c. These results are not quite consistent with the recent DFT+U+SOC study of Hu et al.;[37] using the experimental C2/m structure determined at 300 K, they showed that the ∥(a+c)-direction is the preferred spin direction, in agreement with the recent resonant x-ray magnetic scattering study.[36]

The above-mentioned discrepancy leads us to question whether or not the use of different crystal structures in the DFT+U+SOC calculations is responsible for the discrepancy. Our DFT+U+SOC calculations using the experimental C2/m structure of $Na_2IrO_3$ determined at 300 K are consistent with the results reported by Hu et al.;[37] the ∥(1.5a + 2.3c*)-direction is preferred to the ∥b-direction by 2.77 meV/Ir. This is apparently surprising because one would normally expect the lower-temperature crystal structure to be more reliable in discussing a low-temperature phenomenon such as the spin orientation. To probe the cause for this puzzling observation, we further carry out DFT+U+SOC calculations using the optimized C2/m structure of $Na_2IrO_3$ at 125 K. The 125 K optimized structure predicts that the ∥(1.2a + 2.6c*)-direction is favored over the ∥b-direction (by 5.68 meV/Ir), consequently suggesting that the experimental C2/m structure at 125 K is not accurate for the quantitative determination of the preferred spin orientation. As can be seen from the PDOS plots of **Fig. 5b,c**, the $t_{2g}$-block bands of the optimized and experimental 125 K structures are substantially different (see below). The shapes of the $IrO_6$ octahedra found for the optimized and experimental crystal structures of $Na_2IrO_3$ (determined at both 125 and 300 K) show considerable differences in the lengths of their O…O edges, as summarized in **Table 4S** of the Supporting Material.[48] To correctly determine the preferred spin orientation of $Na_2IrO_3$ by DFT+U+SOC calculations, use of an accurate crystal structure is necessary.

From the quantitative DFT+U+SOC calculations described above, it is clear that the preferred spin orientation of the $Ir^{4+}$ ions in $Na_2IrO_3$ has both ‖c* and ‖a components. Let us now examine the cause for this observation by considering how the inter-site interaction affects relative ordering of the down-spin 1a and 1e states of an $Ir^{4+}$ ion. Consider a dimer made up of two adjacent $Ir^{4+}$ ions and recall that the d-orbital component of the 1a state is the $3z^2$-$r^2$ orbital, and those of the 1e state are the (xy, $x^2$-$y^2$) and (xz, yz) orbitals (**Fig. 2a**). As depicted in **Fig. 5a**, the inter-site interaction between the two 1a states leads to the 1a+ and 1a- states, and that between the 1e states to the 1e+ and 1e- states. The split between 1a+ and 1a- states is weak because the lateral extension of the $3z^2$-$r^2$ orbitals within the plane of the honeycomb layer is small. In contrast, the split between the 1e+ and 1e- states is large because the lateral extension of the (xy, $x^2$-$y^2$) orbitals is large and because so is that of the (xz, yz) orbitals. With four down-spin electrons in the dimer, the 1e- states are empty while the remaining states are filled. The $|\Delta L_z| = 1$ interactions between the 1a+/1a- and 1e- states predict the ⊥z spin orientation. The interactions between the 1e+ and 1e- states give rise to the $|\Delta L_z| = 0$ interactions, between their (xz, yz) sets and between their (xy, $x^2$-$y^2$) sets, predicting the ‖z spin orientation. Consequently, if the 1a+ and 1a- states are close in energy to the 1e+ states, then the preferred spin orientation of the $Ir^{4+}$ ion would be the (⊥z + ‖z) direction. In essence, the ‖a component of the $Ir^{4+}$ spin orientation in $Na_2IrO_3$ is a consequence of the intersite interactions, because only the ‖c* direction is predicted in their absence.

The electronic structure calculated for the optimized 125 K structure is presented in terms of the PDOS plots in **Fig. 5b**, which is well accounted for by the above description; the relative ordering of the down-spin states is best described as $1a_+$, $1a_- \leq 1e_+ < 1e_-$. The PDOS plots for the experimental 125 K structure, presented in **Fig. 5c**, is best described as $1e_+ < 1a_+$, $1a_- < 1e_-$. Then, the $|\Delta L_z| = 1$ interaction can dominate over the $|\Delta L_z| = 0$ interaction thereby predicting the $\perp z$ spin orientation. The latter is consistent with the $\parallel b$ spin orientation computationally found, because the $\parallel b$ direction is perpendicular to the c*-axis direction (i.e., the local z-direction).

## 5. Madelung potentials

As discussed in the previous sections, there are observations that the 5d electrons of $Ir^{4+}$ ions are less localized in $Na_2IrO_3$ than in $Sr_3NiIrO_6$ and $Sr_2IrO_4$. Then, the d-states of each $Ir^{4+}$ ion in $Na_2IrO_3$ would possess information not only about its own local site but also about its surrounding $Ir^{4+}$ ions. The previous section showed that the determination of the $Ir^{4+}$ spin orientation in $Na_2IrO_3$ by DFT+U+SOC calculations is quite sensitive to the accuracy of its crystal structure. This is understandable if the d-states of its $Ir^{4+}$ ions are somewhat delocalized, because the description of their SOC-induced interactions around a given $Ir^{4+}$ site requires the accurate structures around not only its site but also its surrounding $Ir^{4+}$ sites. The local factors affecting electron localization such as the oxidation state and the SOC constant $\lambda$, the $Ir^{4+}$ ions of $Na_2IrO_3$ do not differ from those of $Sr_3NiIrO_6$ and $Sr_2IrO_4$. However, the Ir/O lattice of $Na_2IrO_3$ is surrounded by monovalent cations ($Na^+$), but those of $Sr_3NiIrO_6$ and $Sr_2IrO_4$ by divalent cations ($Ni^{2+}$ and $Sr^{2+}$ in $Sr_3NiIrO_6$, and $Sr^{2+}$ in $Sr_2IrO_4$). Consequently, the Madelung potentials[53] acting on the $Ir^{4+}$ sites of $Na_2IrO_3$ should differ from those of $Sr_3NiIrO_6$ and $Sr_2IrO_4$. The 5d electrons of an $Ir^{4+}$ ion would be more strongly bound (i.e., more strongly localized) to the ion, if its Madelung potential is more attractive (i.e., more negative). To check this possibility, we calculate the Madelung potentials of the $Ir^{4+}$ ions in $Na_2IrO_3$, $Sr_3NiIrO_6$ and $Sr_2IrO_4$ as well as $Nd_3IrO_7$ and $Y_2Ir_2O_7$, which are composed of corner-sharing $IrO_6$ octahedra and in which the $Ir^{4+}$ ions are surrounded with trivalent cations (i.e., $Y^{3+}$ and $Nd^{3+}$). As summarized in **Table 4**, our results show clearly that the Madelung potential of the $Ir^{4+}$ ion is less negative for $Na_2IrO_3$, which has monovalent cations ($Na^+$), than for the remaining iridates, in which the $Ir^{4+}$ ions are surrounded by either divalent cations ($Sr^{2+}$, $Ni^{2+}$) or trivalent cations ($Y^{3+}$ and $Nd^{3+}$). This supports our suggestion that the 5d electrons of the $Ir^{4+}$ are more delocalized in $Na_2IrO_3$ than in $Sr_3NiIrO_6$ and $Sr_2IrO_4$. It is interesting that a non-local factor such as the Madelung potential can influence the extent of electron localization.

## 6. Origin of the spin-half syndrome

It is an undeniable experimental fact that the S = 1/2 ions $Ir^{4+}$ at the octahedral sites of $Sr_3NiIrO_6$, $Sr_2IrO_4$ and $Na_2IrO_3$ possess preferred spin orientations. This is also true for the S = 1/2 ions $Cu^{2+}$ at various square planar sites,[12] and for the S = 1/2 ions $V^{4+}$ at distorted octahedral sites of $R_2V_2O_7$ (R = rare earth).[54,55] For all these S = 1/2 ion cases, DFT calculations reproduce the observed spin orientations if SOC is taken into consideration, and so do the perturbation theory analyses.[12,13] Thus, one must conclude that all these S = 1/2 ions have magnetic anisotropy, and it is caused by the SOC-induced interactions among their crystal-field split d-states. In the DFT and perturbation theory analyses, the energy states of a magnetic system are discussed in terms of its magnetic orbitals (i.e., its singly occupied orbitals). Each magnetic orbital represents either the up-spin state $\left|\uparrow\right\rangle$ or the down-spin state $\left|\downarrow\right\rangle$, so the overall spin S of a magnetic ion (more precisely, its $MO_n$ polyhedron) is related to how many magnetic orbitals the $MO_n$ polyhedron has. Each magnetic ion of a magnetic orbital in spin state $\left|S,S_z\right\rangle$ (i.e., $\left|\uparrow\right\rangle = \left|\frac{1}{2},+\frac{1}{2}\right\rangle$ or $\left|\downarrow\right\rangle = \left|\frac{1}{2},-\frac{1}{2}\right\rangle$) is described by the orbital/spin state $\left|L,L_z\right\rangle\left|S,S_z\right\rangle$. The SOC, $\lambda\hat{S}\cdot\hat{L}$, modifies the magnetic states because it induces intermixing between them, but this intermixing does not occur in the spin part $\left|S, S_z\right\rangle$, but in the orbital part $\left|L, L_z\right\rangle$, of each state. For instance, when there is no degeneracy in the magnetic orbitals, a given magnetic orbital $\left|L, L_z\right\rangle\left|S, S_z\right\rangle$ is modified by the intermixing as

$$\left\{(1-\gamma^2-\delta^2-\cdots)\left|L, L_z\right\rangle + \gamma\left|L', L_z'\right\rangle + \delta\left|L'', L_z''\right\rangle + \cdots \right\}\left|S, S_z\right\rangle,$$

where $\gamma$ and $\delta$ are the mixing coefficients. This SOC-induced orbital mixing is independent of whether the overall spin S of the magnetic ion is 1/2 or greater because it occurs in each individual magnetic orbital and hence does not depend on how many magnetic orbitals a magnetic ion generates. This is why magnetic anisotropy is predicted for S = 1/2 ions on an equal footing to S > 1/2 ions in an electronic Hamiltonian approach. This fundamental result is not described by a spin Hamiltonian simply because it lacks the orbital degree of freedom; having completely suppressed the orbital $\left|L, L_z\right\rangle$ of a magnetic ion, a spin Hamiltonian does not allow one to discuss the SOC, $\lambda\hat{S}\cdot\hat{L}$, and hence is unable to describe the preferred spin orientation of any magnetic ion. The spin-half syndrome is a direct consequence from this deficiency of a spin Hamiltonian.

To a limited extent, a spin Hamiltonian can indirectly discuss the effect of SOC. For example, the phenomenon of spin canting is discussed by introducing the Dzyaloshinskii-Moriya (DM) term, $\vec{D}_{ij}\cdot(\vec{S}_i\times\vec{S}_j)$, between two adjacent spins $\vec{S}_i$ and $\vec{S}_j$.[11,56,57] Here the DM vector $\vec{D}_{ij}$, which is a parameter originating from SOC, is related to the difference in the unquenched orbital angular momenta on the two magnetic sites i and j, namely, $\vec{D}_{ij} = \lambda J_{ij}(\delta\vec{L}_1 - \delta\vec{L}_2)$.[11,57] The DM interaction represents a classic example of the effective spin approximation,[58] in which magnetic ions are treated as spin-only ions by including the effect of SOC associated with their unquenched orbital moments $\delta\vec{L}$ into suitable constants (e.g., the DM constant $\vec{D}_{ij}$ and anisotropic g-factors).[10,57] For a magnetic ion whose HOMO and LUMO are not degenerate, the effective spin approximation reduces the SOC Hamiltonian $\hat{H}_{SO} = \lambda\hat{S}\cdot\hat{L}$ to the zero-field Hamiltonian $\hat{H}_{zf}$ ,[10,12]

$$\hat{H}_{zf} = D\left(\hat{S}_z^2 - \tfrac{1}{3}\hat{S}^2\right) + \tfrac{1}{2}E\left(\hat{S}_+\hat{S}_+ + \hat{S}_-\hat{S}_-\right),\tag{5}$$

with the constants D and E related to the SOC as $D \propto \lambda^2(\delta L_{\parallel} - \delta L_{\perp})$ and $E \propto \lambda^2(\delta L_x - \delta L_y)$. Here $\delta L_{\parallel}$ and $\delta L_{\perp}$ are the unquenched orbital angular momenta along the $\parallel z$ and $\perp z$ directions, respectively, while $\delta L_x$ and $\delta L_y$ are the unquenched orbital angular momenta along the x- and y-directions, respectively. For S > 1/2 ions, Eq. 5 predicts magnetic anisotropy. For instance, a S = 1 ion is described by three spin states, $\left|S,S_z\right\rangle = \left|1,+1\right\rangle$, $\left|1,0\right\rangle$ and $\left|1,-1\right\rangle$. Thus,

$$D\left(\hat{S}_z^2 - \tfrac{1}{3}\hat{S}^2\right)\left|1,+1\right\rangle = D[\hat{S}_z^2 - \tfrac{1}{3}S(S+1)]\left|1,+1\right\rangle = +\tfrac{1}{3}D\left|1,+1\right\rangle$$
$$D\left(\hat{S}_z^2 - \tfrac{1}{3}\hat{S}^2\right)\left|1,0\right\rangle = D[\hat{S}_z^2 - \tfrac{1}{3}S(S+1)]\left|1,0\right\rangle = -\tfrac{2}{3}D\left|1,0\right\rangle$$
$$D\left(\hat{S}_z^2 - \tfrac{1}{3}\hat{S}^2\right)\left|1,-1\right\rangle = D[\hat{S}_z^2 - \tfrac{1}{3}S(S+1)]\left|1,-1\right\rangle = +\tfrac{1}{3}D\left|1,-1\right\rangle$$

and

$$E(\hat{S}_+\hat{S}_+ + \hat{S}_-\hat{S}_-)\left|1,+1\right\rangle = E\left|1,-1\right\rangle$$
$$E(\hat{S}_+\hat{S}_+ + \hat{S}_-\hat{S}_-)\left|1,0\right\rangle = 0$$
$$E(\hat{S}_+\hat{S}_+ + \hat{S}_-\hat{S}_-)\left|1,-1\right\rangle = E\left|1,+1\right\rangle$$

This shows that the $\left|1,\pm1\right\rangle$ states are separated in energy from the $\left|1,0\right\rangle$ state by |D|. In addition, the $\left|1,+1\right\rangle$ and $\left|1,-1\right\rangle$ states interact and become split in energy by |E|. Due to this energy split, the thermal populations of the three states differ, hence leading to magnetic anisotropy.

A rather different situation occurs for a S = 1/2 ion, which is described by two spin states, $\left|\uparrow\right\rangle = \left|\tfrac{1}{2},+\tfrac{1}{2}\right\rangle$ and $\left|\downarrow\right\rangle = \left|\tfrac{1}{2},-\tfrac{1}{2}\right\rangle$. We note that

$$D(\hat{S}_z^2 - \tfrac{1}{3}\hat{S}^2)\left|\tfrac{1}{2},+\tfrac{1}{2}\right\rangle = D[\hat{S}_z^2 - \tfrac{1}{3}S(S+1)]\left|\tfrac{1}{2},+\tfrac{1}{2}\right\rangle = 0$$

$$D(\hat{S}_z^2 - \tfrac{1}{3}\hat{S}^2)\left|\tfrac{1}{2},-\tfrac{1}{2}\right\rangle = D[\hat{S}_z^2 - \tfrac{1}{3}S(S+1)]\left|\tfrac{1}{2},-\tfrac{1}{2}\right\rangle = 0$$

and

$$E(\hat{S}_+\hat{S}_+ + \hat{S}_-\hat{S}_-)\left|\tfrac{1}{2},+\tfrac{1}{2}\right\rangle = 0$$

$$E(\hat{S}_+\hat{S}_+ + \hat{S}_-\hat{S}_-)\left|\tfrac{1}{2},-\tfrac{1}{2}\right\rangle = 0$$

Consequently, the up-spin and down-spin states do not interact under $\hat{H}_{zf}$, so their degeneracy is not split. (This result obeys the Kramers degeneracy theorem,[59] which states that the degeneracy of an odd-spin system should not be split in the absence of an external magnetic field.) This is so even though the constants D and E are nonzero, that is, even though SOC effects are taken into consideration albeit indirectly. Thus, the thermal populations of the two states $\left|\uparrow\right\rangle$ and $\left|\downarrow\right\rangle$ are identical, hence leading to the conclusion that an S = 1/2 ion has no magnetic anisotropy that arise from SOC. This is the origin of the spin-half syndrome.

Note that $\hat{H}_{SO} = \lambda \hat{S} \cdot \hat{L}$ and $\hat{H}_{zf}$ are local (i.e., single-spin site) operators, and do not describe interactions between different spin sites. The SOC-induced magnetic anisotropy for S > 1/2 ions is commonly referred to as the single-ion (or single-site) anisotropy, to which most practitioners of spin Hamiltonian analysis have no objection. However, they deny strenuously that S = 1/2 ions have single-ion anisotropy and suggest the use of the term "magneto-crystalline anisotropy" to describe the experimentally observed magnetic anisotropy of S = 1/2 ions. However, this term is highly misleading, because it implies that the observed anisotropy is not caused by the single-spin site effect (i.e., SOC) but rather by nonlocal effects (i.e., anything other than SOC, e.g., asymmetric spin exchange and magnetic dipole-dipole interactions), just as Moriya and Yoshida argued for the S = 1/2 system $CuCl_2 \cdot 2H_2O$ more than six decades ago.[60] However, as recently shown [12,13] for various magnetic solids of S = 1/2 ions, the spin-half syndrome is erroneous. Unfortunately, this syndrome remains unabated because it has been perpetuated in monographs and textbooks on magnetism.[61-65]

## 7. Discussion

The energy stabilization $\Delta E$ associated with the SOC-induced interaction between the HOMO and the LUMO (with energies $e_{HO}$ and $e_{LU}$, respectively) is given by Eq. 6.[11,13]

$$\Delta E = \begin{cases} -\left| \lambda \langle \psi_{HO} | \hat{S} \cdot \hat{L} | \psi_{LU} \rangle \right|, & \text{if } e_{HO} = e_{LU} \\ -\dfrac{\lambda^2 \left| \langle \psi_{HO} | \hat{S} \cdot \hat{L} | \psi_{LU} \rangle \right|^2}{\left| e_{HO} - e_{LU} \right|}, & \text{if } e_{HO} < e_{LU} \end{cases} \quad (6)$$

where $\left| e_{HO} - e_{LU} \right|$ is the HOMO-LUMO energy difference. The overall widths of the $t_{2g}$-block bandwidths in $Sr_3NiIrO_6$, $Sr_2IrO_4$ and $Na_2IrO_3$ are of the order of 2 eV (i.e., 1.7, 2.6 and 2.4 eV, respectively from our DFT+U calculations) and the $\left| e_{HO} - e_{LU} \right|$ values are of the order of 0.2 eV (0.2, 0.2 and 0.3 eV, respectively, see the PDOS plots of **Fig. 2**, **4b** and **5a**). The SOC constant $\lambda$ of $Ir^{4+}$ is of the order of 0.5 eV [65] so that $\lambda^2$ is comparable in magnitude to $\left| e_{HO} - e_{LU} \right|$ for the case of $e_{HO} < e_{LU}$. In such a case, use of perturbation theory does not lead to an accurate estimation of $\Delta E$. However, thisdoes not affect our qualitative predictions of the preferred spin orientations, because the latter do not require a quantitative evaluation of $\Delta E$.

The effects of SOC are discussed in terms of either the LS or the jj coupling scheme depending on the strength of SOC. In the LS (or Russel-Saunders) scheme the electron spin momenta are summed up to find the total spin momentum $\vec{S} = \sum \vec{s}_i$, and the orbital momenta of individual electrons to find the total orbital momentum $\vec{L} = \sum \vec{l}_i$. Then, the SOC is included to couple $\vec{S}$ and $\vec{L}$ to obtain the total angular momentum $\vec{J}$, leading to the SOC Hamiltonian, $\hat{H}_{SO} = \lambda \vec{S} \cdot \vec{L}$. The LS coupling scheme is typically employed for elements with weak SOC (e.g., 3d- and 4d-elements). In this scheme the crystal-field split $d$-states of a $M$O$_n$ polyhedron are closely related to the orbital states $\left| L, L_z \right\rangle$ of $M$ in the up-spin $\left| \uparrow \right\rangle$ or down-spin state $\left| \downarrow \right\rangle$ magnetic orbitals of $M$O$_n$. As discussed for Sr$_3$NiIrO$_6$, Sr$_2$IrO$_4$ and Na$_2$IrO$_3$ in Sections 3 and 4, our analyses based on the LS coupling scheme explain the spin-orbit Mott insulating states of these 5d oxides as well as their observed magnetic anisotropies. The jj coupling scheme, appropriate for elements with strong SOC (e.g., 4f and 5f elements), has recently become popular in discussing the spin-orbit Mott insulating states of 5d oxides.[16] In this scheme, the spin and orbital momenta are added to obtain the total angular momentum $\vec{j}_i = \vec{l}_i + \vec{s}_i$ for each electron of a magnetic ion $M$, and the $\vec{j}_i$'s of the individual electrons are added to find the total angular momentum, $\vec{J} = \sum \vec{j}_i$, of $M$. In this approach, it is not obvious how to relate the $\vec{J}$ states to the crystal-field split $d$-states of $M$O$_n$ unless the corresponding analysis is done by using the LS coupling scheme, because the crystal-field split d-states of $M$O$_n$ are determined by the interactions of the orbital states $\left| L, L_z \right\rangle$ of $M$ with the 2p orbitals of the surrounding O ligands and because the information about the orbital states $\left| L, L_z \right\rangle$ of $M$ is completely hidden in the jj coupling scheme. As a consequence, use of the jj scheme makes it difficult to predict such fundamental magnetic properties as the preferred spin orientation and the uniaxial magnetism of a magnetic ion $M$. The latter are readily predicted by the LS coupling scheme. As found for the Ir$^{4+}$ ion of Sr$_3$NiIrO$_6$, the need to employ "J-states" in the LS scheme arises only when a magnetic ion has an unevenly-filled degenerate d-state, leading to an unquenched orbital momentum $\vec{L}$ that combines with $\vec{S}$ to form $\vec{J} = \vec{S} + \vec{L}$. In the LS scheme, use of J-states is not appropriate for $Sr_2IrO_4$ and $Na_2IrO_3$ because they possess no unquenched orbital momentum $\vec{L}$ to combine with $\vec{S}$. Our discussion shows that the magnetic properties of the 5d oxides are better explained by the LS scheme than by the jj scheme, and hence implies that the spin-orbital entanglement in 5d elements, though stronger than those in 3d and 4d elements, may not be as strong as has been put forward.[16] This conclusion is consistent with the view that SOC for 5d elements lies in between the LS and jj coupling schemes, but is closer to the LS scheme.[66]

## 8. Concluding remarks

The S = 1/2 ions $Ir^{4+}$ at the octahedral sites of $Sr_3NiIrO_6$, $Sr_2IrO_4$ and $Na_2IrO_3$ exhibit preferred spin orientations in coordinate space because SOC induces interactions among the crystal-field split d-states of their $IrO_6$ octahedra and because the associated energy-lowering depends on the spin orientation. The preferred spin orientations of the $Ir^{4+}$ ions are predicted by considering the SOC-induced HOMO-LUMO interactions of their $IrO_6$ octahedra present in crystalline oxides. The $Ir^{4+}$ spin of $Na_2IrO_3$ has nonzero components along the $\|z$ and $\perp z$ directions (with respect to the pseudo 3-fold rotational axis) because the intersite interactions are substantial. With the z-direction of each $IrO_6$ octahedron along the axis of its highest rotational symmetry, the $Ir^{4+}$ spins of $Sr_3NiIrO_6$ and $Sr_2IrO_4$ exhibit either the $\|z$ or the $\perp z$ direction because the crystal-field split d-states of their $IrO_6$ are not as strongly affected by the intersite interactions as those of the $IrO_6$ in $Na_2IrO_3$. The 5d electrons of $Na_2IrO_3$ are less localized than are those of $Sr_3NiIrO_6$ and $Sr_2IrO_4$, most probably because the Madelung potentials of the $Ir^{4+}$ ions are less negative for $Na_2IrO_3$ than for $Sr_3NiIrO_6$ and $Sr_2IrO_4$. In both the DFT and perturbation theory analyses, only the orbital parts of electronic states are modified by SOC so that S = 1/2 ions are predicted to possess magnetic anisotropy, in agreement with experiment. The spin-half syndrome results from the limited nature of a spin Hamiltonian that it lacks the orbital degree of freedom. The magnetic anisotropy of the $Ir^{4+}$ ions in $Sr_3NiIrO_6$, $Sr_2IrO_4$ and $Na_2IrO_3$ is well explained by the LS coupling scheme, but not by the jj scheme.

**Acknowledgments**

This research used resources of the National Energy Research Scientific Computing Center, a DOE Office of Science User Facility supported by the Office of Science of the U.S. Department of Energy under Contract No. DE-AC02-05CH11231. MHW thanks Prof. Sang-Wook Cheong and Dr. Jae Wook Kim for sharing their experimental data on $Sr_3NiIrO_6$ and $Sr_3NiPtO_6$ in the initial development stage of this work.

Table 1. The relative energies $\Delta E = E_{AFM} - E_{FM}$ between the AFM and FM states (in meV/FU) obtained from DFT+U calculations

| $U_{eff}$ = 2 eV for Ir | | $U_{eff}$ = 4 eV for Ni | |
|---|---|---|---|
| $U_{eff}$ for Ni (eV) | $\Delta E$ (meV/FU) | $U_{eff}$ for Ir (eV) | $\Delta E$ (meV/FU) |
| 4 | -24 | 0 | -58 |
| 5 | -37 | 1 | -69 |
| 6 | -50 | 2 | -24 |

Table 2. Results of DFT+U+SOC calculations for $Sr_3NiIrO_6$ using $U_{eff} = 4 – 6$ eV for Ni with $U_{eff}$ fixed at 2 eV for Ir.

(a) Relative energies (in meV/FU) of the ∥c and ⊥c spin orientations for cases adjacent $Ni^{2+}$ and $Ir^{4+}$ spins in each $NiIrO_6$ chains have FM and AFM couplings.

|  | Spin orientation | $U_{eff} = 4$ eV | $U_{eff} = 5$ eV | $U_{eff} = 6$ eV |
|---|---|---|---|---|
| FM | ∥c | 0.00 | 0.00 | 0.00 |
|  | ⊥c | -3.47 | -5.37 | -7.13 |
| AFM | ∥c | 0.00 | 0.00 | 0.00 |
|  | ⊥c | 11.6 | 11.1 | 10.7 |

(b) Orbital moment $\mu_L$ and spin moment $\mu_S$ (in units of $\mu_B$) of the $Ni^{2+}$ and $Ir^{4+}$ ions for cases when adjacent $Ni^{2+}$ and $Ir^{4+}$ spins in each $NiIrO_6$ chains have FM and AFM couplings with their spins aligned along the ∥c direction.

|  |  | $U_{eff} = 4$ eV | | $U_{eff} = 5$ eV | | $U_{eff} = 6$ eV | |
|---|---|---|---|---|---|---|---|
|  |  | Ir | Ni | Ir | Ni | Ir | Ni |
| FM | $\mu_S$ | 0.61 | 1.66 | 0.60 | 1.71 | 0.60 | 1.76 |
|  | $\mu_L$ | -0.07 | 0.05 | -0.07 | 0.04 | -0.08 | 0.04 |
| AFM | $\mu_S$ | -0.58 | 1.66 | -0.60 | 1.71 | -0.60 | 1.76 |
|  | $\mu_L$ | -0.34 | 0.05 | -0.34 | 0.05 | -0.34 | 0.04 |

Table 3. Results of DFT+U+SOC calculations for the axially-elongated and axially-compressed structures of $Sr_2IrO_4$ using $U_{eff} = 2$ eV for Ir. Except for the spin orientations, the arrangement of the $Ir^{4+}$ spins is identical to the experimentally observed one.

(a) The relative energies $\Delta E$ (meV/FU) of the ∥a, ∥b and ∥c spin orientations.

| Spin orientation | Axially-elongated | Axially-compressed |
|---|---|---|
| ∥a | 0.00 | 6.45 |
| ∥b | 0.02 | 6.65 |
| ∥c | 1.81 | 0.00 |

(b) The orbital moment $\mu_L$ and spin moment $\mu_S$ (in units of $\mu_B$) of the $Ir^{4+}$ ions.

| | Axially-elongated | | | Axially-compressed | | |
|---|---|---|---|---|---|---|
| | ∥a | ∥b | ∥c | ∥a | ∥b | ∥c |
| $\mu_S$ | 0.13 | 0.13 | 0.29 | 0.03 | 0.03 | 0.57 |
| $\mu_L$ | 0.30 | 0.30 | 0.40 | 0.27 | 0.27 | 0.50 |

Table 4. Madelung potentials[50] ($\text{Å}^{-1}$) at the $Ir^{4+}$ sites in some iridates

| | Madelung potential | Structural feature |
|---|---|---|
| $Nd_3IrO_7$ | -3.56867 | $IrO_5$ chains of corner-sharing $IrO_6$ octahedra |
| $Y_2Ir_2O_7$ | -3.07575 | Pyrochlore lattice of corner-sharing $IrO_6$ octahedra |
| $Sr_2IrO_4$ | -2.94508 | $IrO_4$ layers of corner-sharing $IrO_6$ octahedra |
| $Sr_3NiIrO_6$ | -2.89692 | Face-sharing $IrO_6$ octahedra and $NiO_6$ trigonal prisms |
| $Na_2IrO_3$ | -2.76739 | Honeycomb layer of edge-sharing $IrO_6$ octahedra |

**Figure captions**

Figure 1. (a) A perspective view of an isolated $NiIrO_6$ chain of $Sr_3NiIrO_6$, running along the c-direction, where Ni = blue circle, Ir = red circle, and O = white circle. (b) A projection view of the crystal structure of $Sr_3NiIrO_6$ along the c-direction, where Sr = yellow circle. (c) A view of an isolated $Sr_2IrO_4$ layer made up of corner-sharing axially-elongated $IrO_6$ octahedra approximately along the c-direction. (d) A perspective view of how the $Sr_2IrO_4$ layers stack along the c-direction. (e) A projection view of a $NaIrO_3$ honeycomb layer made up of edge-sharing $IrO_6$ octahedra with Na (light blue circle) at the center of each $Ir_6$ hexagon. (f) A perspective view of how the honeycomb $NaIrO_3$ layers repeat along the c-direction in $Na_2IrO_3$, where the layer of Na atoms lying in between the $NaIrO_3$ honeycomb layers is not shown for simplicity.

Figure 2. (a) The d-orbital compositions of the $t_{2g}$ state of an $IrO_6$ octahedron in case when the local z-axis is taken along the 3-fold rotational axis. (b, c) The two different electron configurations of a low-spin $Ir^{4+}$ ($S = 1/2$, $d^5$) ion at an octahedral site. In this spin-polarized description, the energy separation between the up-spin and down-spin states are exaggerated. (d, e) The PDOS plots for the down-spin d-states of $Ir^{4+}$ in $Sr_3NiIrO_6$ in cases when adjacent $Ni^{2+}$ ($S = 1$) and $Ir^{4+}$ ($S = 1/2$) ions in each $NiIrO_6$ chain are antiferromagnetically coupled in (d), and ferromagnetically coupled in (e). These PDOS plots obtained from DFT+U calculations with $U_{eff} = 4$ eV for Ni and $U_{eff} = 2$ eV for Ir. The numbers (2, -2), (1, -1) and 0 refer respectively to ($L_z = 2$, $L_z = -2$), ($L_z = 1$, $L_z = -1$) and $L_z = 0$, namely, the (xy, $x^2$-$y^2$), (xz, yz) and $3z^2$-$r^2$ sets.

Figure 3. The spin-polarized $(t_{2g})^5$ configurations of the $Ir^{4+}$ ion in (a) the axially-elongated $IrO_6$ octahedron along the 4-fold rotational axis, (b) the axially-compressed $IrO_6$ octahedron along the 4-fold rotational axis, and (c) the axially-compressed $IrO_6$ octahedron along the pseudo 3-fold rotational axis. The local z-axis is taken along the 4-fold rotational axis in (a) and (b), and along the pseudo 3-fold rotational axis in (c).

Figure 4. The d-states of $Sr_2IrO_4$: (a, b) The PDOS plots for the d-states of $Ir^{4+}$ in $Sr_2IrO_4$ in cases when the $IrO_6$ octahedra are axially elongated in (a), and axially compressed in (b). These PDOS plots were obtained from DFT+U calculations with $U_{eff} = 2$ eV. (c) The interaction between adjacent xz orbitals (or adjacent yz orbitals) through the O 2p orbitals through each bent $Ir-O_{eq}-O$ bridge. (d) The interaction between adjacent xy orbitals through the O 2p orbitals through each bent $Ir-O_{eq}-O$ bridge. (e, f) The split d-states of a dimer made up of two adjacent $Ir^{4+}$ ions after incorporating the effect of the inter-site interactions for the cases of the axially-elongated $IrO_6$ octahedra in (e) and the axially-compressed $IrO_6$ octahedra in (f).

Figure 5. The d-states of $Na_2IrO_3$: (a) The split d-states of a dimer made up of two adjacent $Ir^{4+}$ ions after incorporating the effect of the inter-site interactions. (b, c) The PDOS plots calculated of the d-states of calculated for the optimized 125 K structure of $Na_2IrO_3$ in (b), and the experimental 125 K structure of $Na_2IrO_3$ in (c).

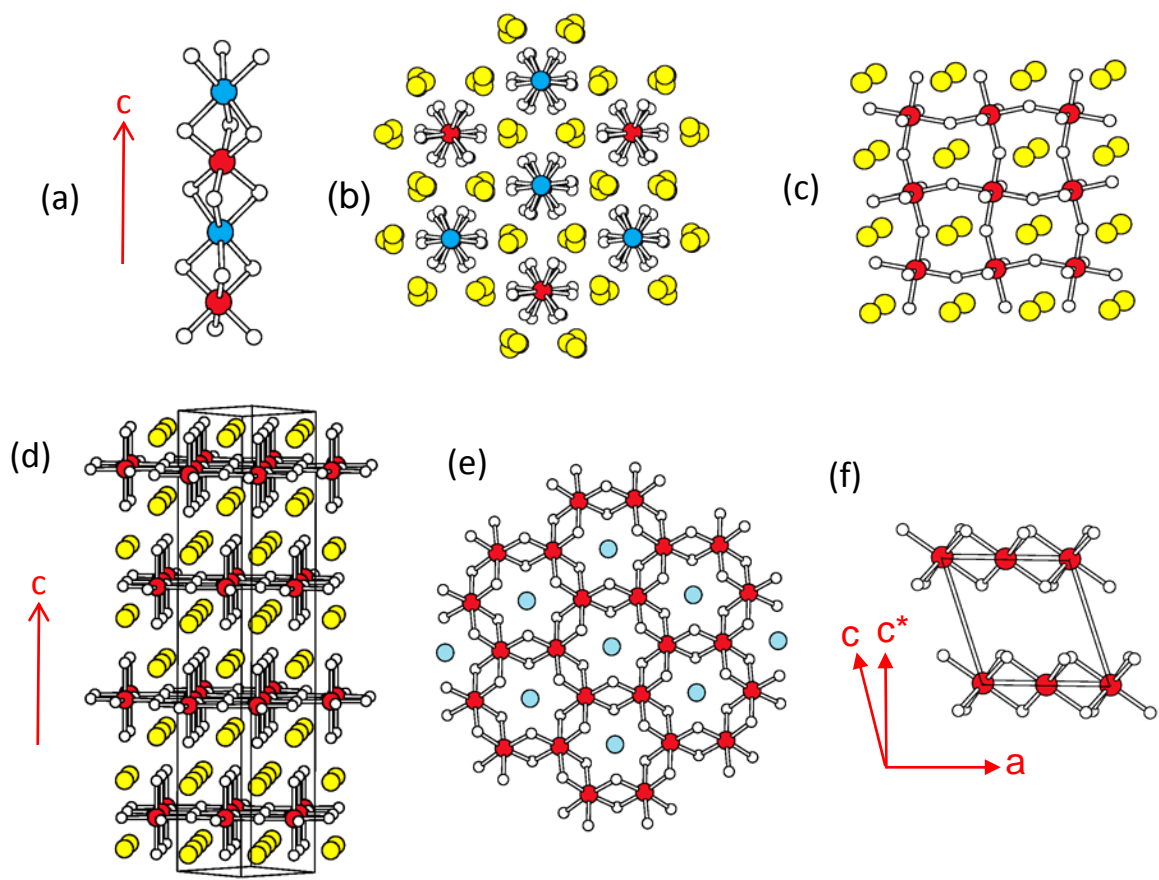

Fig. 1

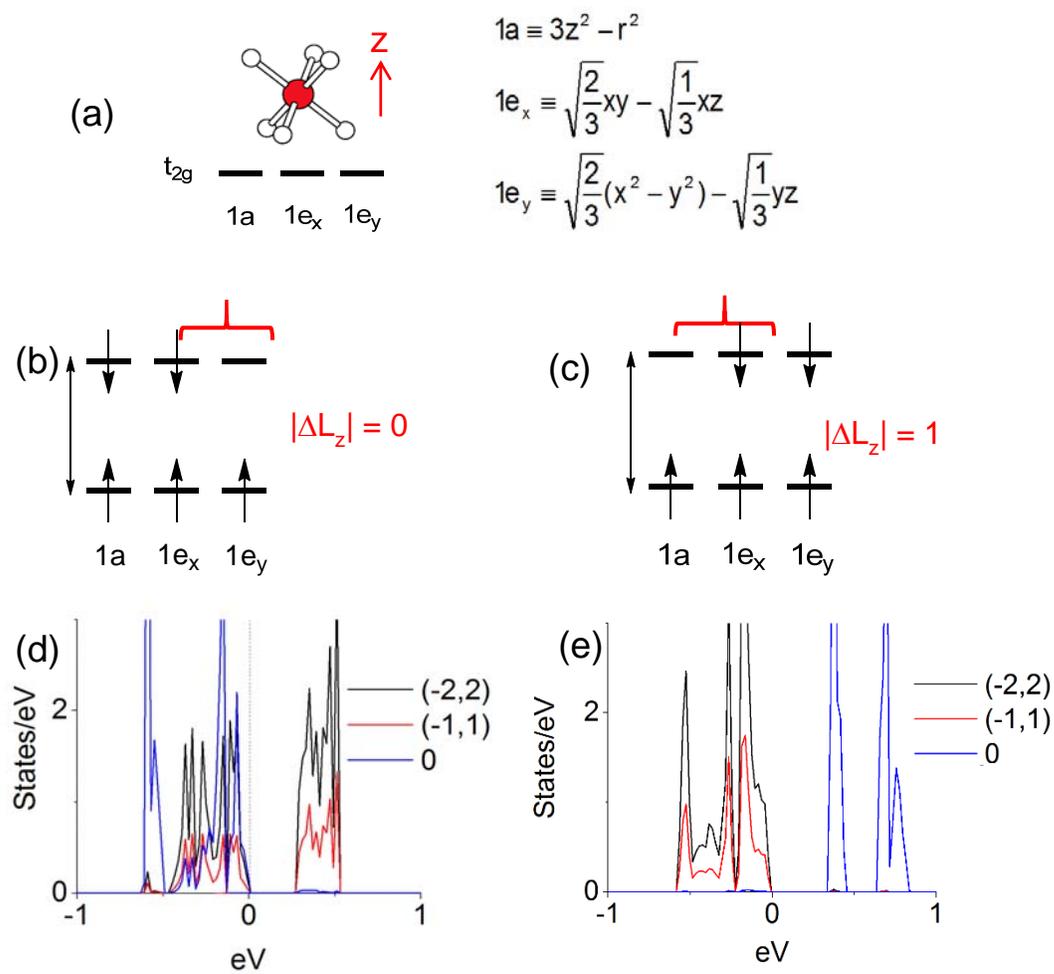

Fig. 2

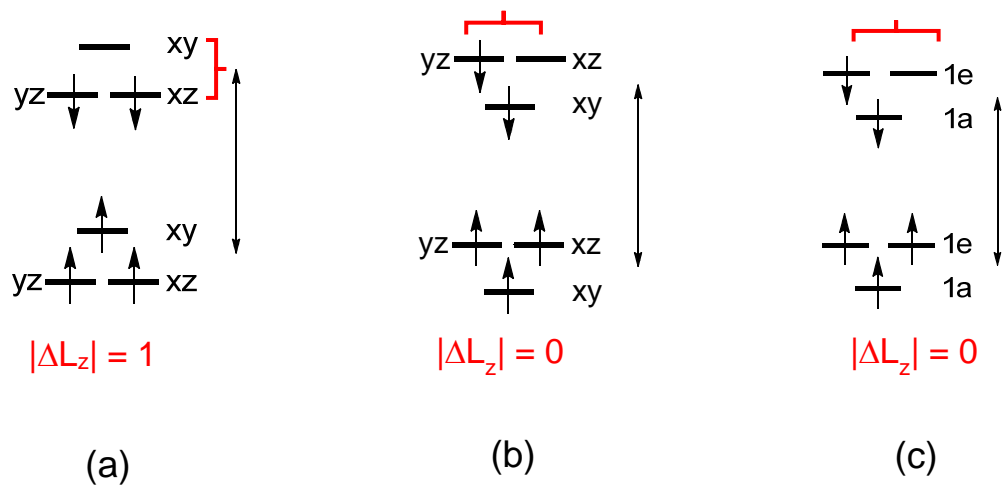

(a)        (b)       (c)

Fig. 3

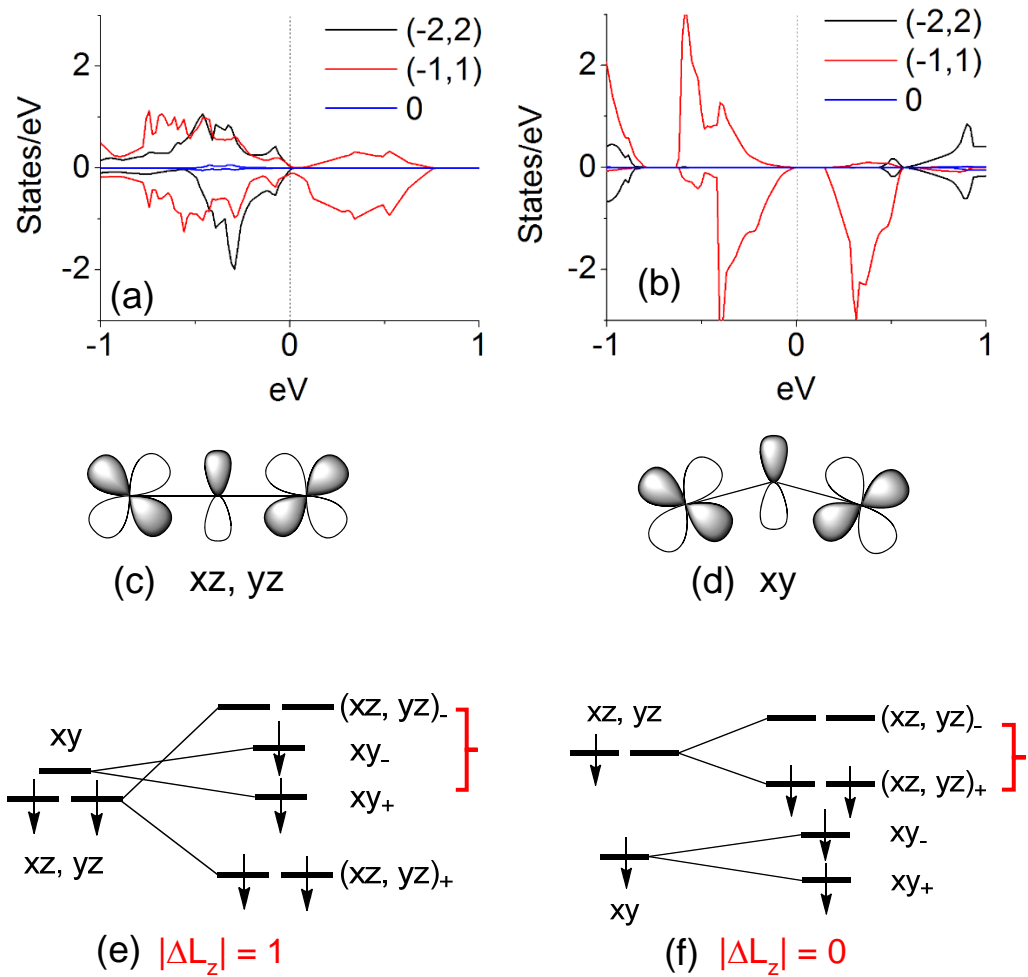

Fig. 4

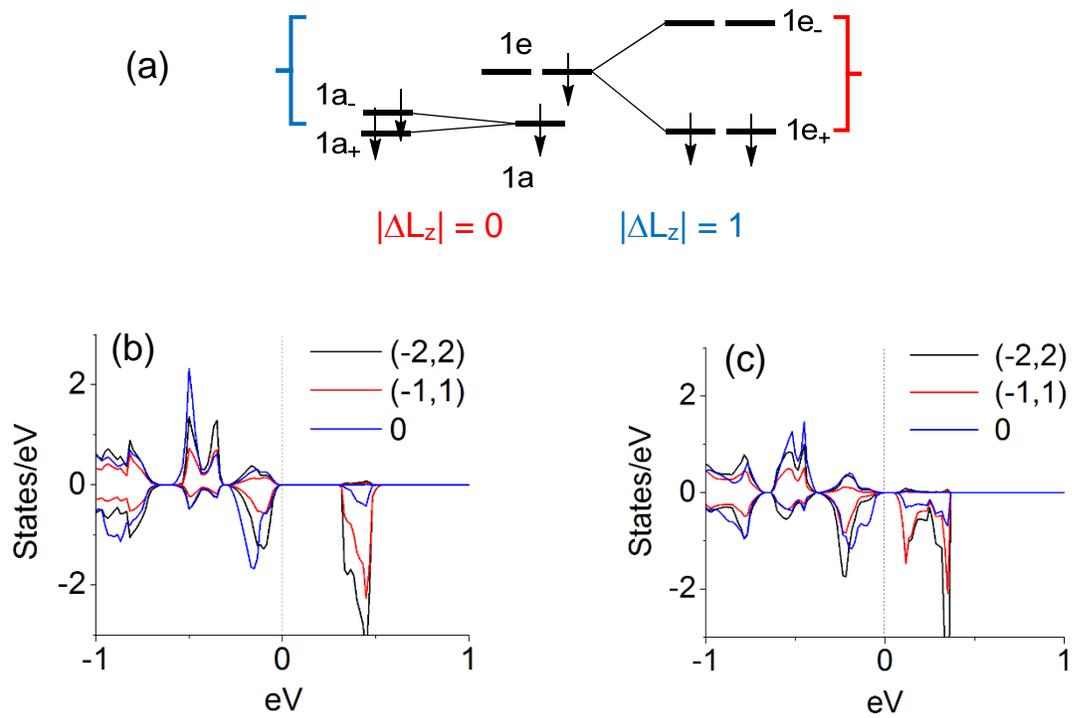

Fig. 5

Supplemental Material

for

**Spin orientations of the spin-half $Ir^{4+}$ ions in $Sr_3NiIrO_6$, $Sr_2IrO_4$ and $Na_2IrO_3$: Density functional, perturbation theory and Madelung potential analyses**

Elijah E. Gordon[1], Hongjun Xiang[2,3], Jürgen Köhler[4], and Myung-Hwan Whangbo[1,*]

[1] Department of Chemistry, North Carolina State University, Raleigh, NC 27695-8204, USA

[2] Key Laboratory of Computational Physical Sciences (Ministry of Education), State Key Laboratory of Surface Physics, Collaborative Innovation Center of Advanced Microstructures, and Department of Physics, Fudan University, Shanghai 200433, P. R. China

[3] Collaborative Innovation Center of Advanced Microstructures, Nanjing 210093, P. R. China

[4] Max-Planck-Institut für Festkörperforschung, D-70569 Stuttgart, Germany

# 1. Effect of the SOC and weak electron correlation on the $Ir^{4+}$ ions of $Sr_3NiIrO_6$ in creating a band gap

We have determined the band gap of $Sr_3NiIrO_6$ by performing DFT+U and DFT+U+SOC calculations with $U_{eff}(Ni)$ fixed at 4 eV but varying $U_{eff}(Ir)$ from 0 to 2 eV. We considered the ferromagnetic (FM) and antiferromagnetic (AFM) arrangements between adjacent $Ni^{2+}$ and $Ir^{4+}$ spins in each $NiIrO_6$ chain. In our DFT+U+SOC calculations, the ∥c and ⊥c spin orientations are also considered for the FM and AFM states, to find that the ⊥c spin orientation is more stable in the FM arrangement but the ∥c spin orientation is for the AFM arrangement. Results of our band gap calculations are summarized in **Table S1**. For $U_{eff}(Ir) = 0$, there is no band gap either in the DFT+U calculations, and very small band gaps in the DFT+U+SOC calculations (namely, 5 and 3 meV for the FM and AFM states, respectively). That is, the SOC of Ir alone cannot induce a substantial band gap. For $U_{eff}(Ir) = 1$ eV, the DFT+U calculations show a larger band gap for the FM than for the AFM arrangement (8 meV vs. 0). The DFT+U+SOC calculations induce a band gap for both FM and AFM arrangements (134 vs. 8 meV, respectively). For $U_{eff}(Ir) = 2$ eV, the band gaps of the FM and AFM arrangements are 365 and 270 meV, respectively, from the DFT+U calculations, and 395 and 240 meV, respectively, from the DFT+U+SOC calculations. It is clear from these results that the cooperation of the SOC and weak electron correlation is essential in creating a band gap in $Sr_3NiIrO_6$, as found for $Ba_2NaOsO_6$ [1].

Table S1. The band gaps (in meV) of $Sr_3NiIrO_6$ calculated for the FM and AFM arrangements of adjacent $Ni^{2+}$ and $Ir^{4+}$ spins in the $NiIrO_6$ chains by DFT+U and DFT+U+SOC calculations with $U_{eff}(Ni) = 4$ eV as a function of $U_{eff}(Ir)$ (= 0, 1, 2 eV).[a]

| | | $U_{eff}(Ir) = 0$ | $U_{eff}(Ir) = 1$ eV | $U_{eff}(Ir) = 2$ eV |
|---|---|---|---|---|
| DFT+U | FM | 0 | 8 | 365 |
| | AFM | 0 | 0 | 270 |
| DFT+U+SOC | FM (spins along $\perp$c)[b] | 5 | 134 | 395 |
| | AFM (spins along ‖c)[b] | 3 | 8 | 240 |

[a] The AFM arrangement is more stable than the FM arrangement for all DFT+U calculations, namely, 58, 69 and 24 meV per FU for the $U_{eff}(Ir) = 0$, 1 and 2 eV, respectively.

[b] In the DFT+U+SOC calculations, the FM arrangement is more stable for the $\perp$c spin orientation than for the ‖c spin orientation, but the opposite is the case for the AFM arrangement.

## 2. The optimized atom positions of the C2/m crystal structures of $Na_2IrO_3$ at 125 K by DFT+U calculations

Table S2. The atom positions in the optimized 125 K structure of $Na_2IrO_3$ ($a$ = 5.319 Å, $b$ = 9.215 Å, $c$ = 5.536 Å, and $\beta$ = 108.67°)[a]

| Atom | x | y | z |
|------|-----------|-----------|-----------|
| Ir1 | 0.000026 | 0.333195 | 0.000006 |
| Ir2 | -0.000026 | 0.666805 | -0.000006 |
| Ir3 | 0.499981 | 0.833182 | -0.000007 |
| Ir4 | 0.500017 | 0.166818 | 0.000004 |
| Na1 | 0.000002 | -0.000000 | 0.000001 |
| Na2 | 0.500002 | 0.500000 | 0.000001 |
| Na3 | -0.000005 | 0.841449 | 0.500012 |
| Na4 | 0.000008 | 0.158551 | 0.499987 |
| Na5 | 0.500010 | 0.341449 | 0.499992 |
| Na6 | 0.499992 | 0.658551 | 0.500006 |
| Na7 | -0.000000 | 0.500000 | 0.500001 |
| Na8 | 0.499999 | -0.000000 | 0.500001 |
| O1 | 0.241073 | 0.319558 | 0.783827 |
| O2 | 0.758929 | 0.680442 | 0.216176 |
| O3 | 0.758935 | 0.319577 | 0.216206 |
| O4 | 0.241066 | 0.680424 | 0.783796 |
| O5 | 0.741048 | 0.819579 | 0.783801 |
| O6 | 0.258952 | 0.180420 | 0.216202 |
| O7 | 0.258915 | 0.819557 | 0.216182 |
| O8 | 0.741086 | 0.180443 | 0.783820 |
| O9 | 0.287594 | -0.000007 | 0.786646 |
| O10 | 0.712399 | 0.000008 | 0.213350 |
| O11 | 0.787601 | 0.500027 | 0.786651 |
| O12 | 0.212396 | 0.499973 | 0.213347 |

[a] The optimization was carried out by using the space group, P1. The optimized results are described by the space group, C2/m.

Table S3. The atom positions in the optimized 300 K structure of $Na_2IrO_3$ ($a$ = 5.427 Å, $b$ = 9.395 Å, $c$ = 5.614 Å, and $\beta$ = 109.04°)[a]

| Atom | x | y | z |
|------|-----------|-----------|-----------|
| Ir1  | 0.502796  | 0.166127  | 0.000003  |
| Ir2  | 0.497201  | 0.833873  | -0.000002 |
| Ir3  | 0.002696  | 0.666126  | -0.000135 |
| Ir4  | -0.002697 | 0.333874  | 0.000138  |
| Na1  | -0.000000 | 0.000000  | -0.000002 |
| Na2  | 0.499999  | 0.500000  | 0.000001  |
| Na3  | 0.500004  | 0.000000  | 0.500001  |
| Na4  | -0.000001 | 0.499999  | 0.500001  |
| Na5  | 0.497212  | 0.339265  | 0.497328  |
| Na6  | 0.502789  | 0.660735  | 0.502671  |
| Na7  | -0.002851 | 0.839608  | 0.497435  |
| Na8  | 0.002861  | 0.160389  | 0.502557  |
| O1   | 0.738586  | 0.180153  | 0.790515  |
| O2   | 0.261412  | 0.819848  | 0.209485  |
| O3   | 0.252232  | 0.181099  | 0.210651  |
| O4   | 0.747767  | 0.818901  | 0.789349  |
| O5   | 0.238433  | 0.680236  | 0.790224  |
| O6   | 0.761565  | 0.319765  | 0.209778  |
| O7   | 0.752503  | 0.681442  | 0.210579  |
| O8   | 0.247495  | 0.318558  | 0.789423  |
| O9   | 0.707049  | 0.005387  | 0.210838  |
| O10  | 0.292952  | -0.005387 | 0.789160  |
| O11  | 0.207249  | 0.505368  | 0.210797  |
| O12  | 0.792749  | 0.494633  | 0.789204  |

[a] The optimization was carried out by using the space group, P1. The optimized results are described by the space group, $P\bar{1}$ .

## 3. Comparison of the structural parameters of the IrO$_6$ octahedra in the experimental and optimized structures at 300 and 125 K

The extent of the axial compression in each distorted IrO$_6$ octahedron can be estimated by the average of the six ∠O-Ir-O angles from the upper and lower IrO$_3$ pyramids of an IrO$_6$ octahedron defined by its pseudo 3-fold rotational axis along the c*-direction (**Fig. S1**). The extent of the inter-site interaction is determined by the nearest-neighbor Ir···Ir distances of the honeycomb layers and also by how symmetrical each distorted IrO$_6$ octahedron is. The latter can be evaluated by inspecting the Ir−O bond lengths as well as the six nearest-neighbor O···O contacts between the upper and lower IrO$_3$ pyramids of an IrO$_6$ octahedron. In **Fig. S1**, the O···O edges labeled as *a*, *b* and *c* are involved in the edge-sharing with adjacent IrO$_6$ octahedra, while those labeled as *a′*, *b′* and *c′* are not. **Table S4** summarizes the average ∠O-Ir-O angles, the nearest-neighbor Ir···Ir distances, the Ir−O bond lengths and the nearest-neighbor O···O distances of the four Cm/2 structures. The experimental crystal structures show that the axial compression is slightly greater for the 125 K than for the 300 K structure (∠O-Ir-O = 94.7 vs. 94.0° in average), and that the inter-site interaction is greater for the 125 K than for the 300 K structure (Ir···Ir = 3.07 vs. 3.18 Å in average). In agreement with the experimental trend, the optimized crystal structures show that the inter-site interaction is greater for the 125 K than for the 300 K structure (Ir···Ir = 3.07 vs. 3.13 Å in average). In contrast to the experimental trend, however, the optimized structures reveal that the axial compression is slightly weaker for the 125 K than for the 300 K structure (∠O-Ir-O = 92.4 vs. 93.4° in average). The O···O edges *a*, *b* and *c* are considerably shorter the O···O edges *a′*, *b′* and *c′* in the experimental 300 K structure as well as in the optimized 300 and 125 K structures. In contrast, however, the difference between the two sets of O···O edges is considerably smaller for the experimental 125 K structure. Furthermore, the variation in the Ir−O bond lengths of each $IrO_6$ octahedron is least in the experimental 125 K structure.

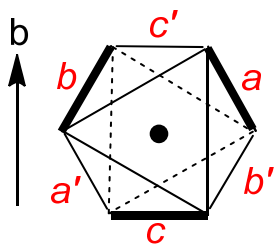

Fig. S1. A projection view of a distorted $IrO_6$ octahedron in $Na_2IrO_3$ along the c*-direction, which is the pseudo 3-fold rotational axis of the octahedron. The Ir atom is represented by a black circle at the center. The solid and dashed triangles represent the triangular bases of the two $IrO_3$ pyramids. The six O···O edges of the $IrO_6$ octahedron are divided into two sets, the three O···O edges represented by thick lines edge-share with neighboring $IrO_6$ octahedra, while the remaining three represented by thin lines do not.

Table S4. The geometrical parameters of the Cm/2 structures of $Na_2IrO_3$ determined at 300 and 125 K as well as the corresponding ones from the structures optimized by DFT+U calculations.

| | 125 K [a] | | 300 K [b] | |
|---|---|---|---|---|
| | Exptl | Optimized | Exptl | Optimized |
| ∠ O-Ir-O (°) | 94.7 | 92.4 | 94.0 | 93.4 |
| Ir⋯Ir (Å) | 3.074 (×2) <br> 3.070 (×4) | 3.074 (×2) <br> 3.070 (×4) | 3.138 (×2) <br> 3.130 (×4) | 3.164 (×2) <br> 3.112 (×4) |
| O⋯O (Å) | a = b = 2.76 <br> a′ = b′ = 2.83 <br> c = 2.77 <br> c′ = 2.85 | a = b = 2.69 <br> a′ = b′ = 2.95 <br> c = 2.70 <br> c′ = 3.02 | a = 2.66 <br> a′ = 3.01 <br> b = 2.69 <br> b′ = 2.93 <br> c = 2.68 <br> c′ = 3.01 | a = b = 2.69 <br> a′ = b′ = 2.93 <br> c = 2.66 <br> c′ = 2.98 |
| Ir–O (Å) | 2.06 (×2) <br> 2.07 (×4) | 2.02 (×2) <br> 2.04 (×2) <br> 2.06 (×2) | 2.06 (×2) <br> 2.07 (×2) <br> 2.08 (×2) | 2.01 (×2) <br> 2.08 (×2) <br> 2.09 <br> 2.10 |

[a] $a$ = 5.319 Å, $b$ = 9.215 Å, $c$ = 5.536 Å, and $\beta$ = 108.67°

[b] $a$ = 5.427 Å, $b$ = 9.395 Å, $c$ = 5.614 Å, and $\beta$ = 109.04°